\UseRawInputEncoding

 \documentclass[twocolumn]{aastex63}

\usepackage{graphicx}
\usepackage{rotating}

\def\msun{\hbox{M$_\odot$}}

\def\t4{\hbox{t$_{\rm 4}$}}

\def\msunyr{\hbox{M$_\odot$yr$^{-1}$}}

\def\cm3{\hbox{cm$^{-3}$}}
\def\one{\,{\sc i}}             
\def\two{\,{\sc ii}}
\def\three{\,{\sc iii}}
\def\four{\,{\sc iv}}
\received{June 1, 2019}
\revised{January 10, 2019}
\accepted{\today}
\submitjournal{ApJ}

\shorttitle{CLUES}
\shortauthors{Sirressi et al.}
\graphicspath{{./}{figures/}}

\begin{document}

\title{CLusters in the Uv as EngineS (CLUES): I. Survey presentation \& FUV spectral analysis of the stellar light}

\correspondingauthor{Mattia Sirressi}
\email{mattia.sirressi@astro.su.se}

\author{Mattia Sirressi}
\affiliation{Department of Astronomy, Oskar Klein Centre, Stockholm University, AlbaNova University
Centre, SE-106 91 Stockholm, Sweden}

\author[0000-0002-0786-7307]{Angela Adamo}
\affiliation{Department of Astronomy, Oskar Klein Centre, Stockholm University, AlbaNova University
Centre, SE-106 91 Stockholm, Sweden}

\author{Matthew Hayes}
\affiliation{Department of Astronomy, Oskar Klein Centre, Stockholm University, AlbaNova University
Centre, SE-106 91 Stockholm, Sweden}

\author{Shannon Osborne}
\affiliation{Space Telescope Science Institute, 3700 San Martin Drive, Baltimore, MD 2121, USA}

\author[0000-0003-4857-8699]{Svea Hernandez}
\affiliation{AURA for ESA, Space Telescope Science Institute, 3700 San Martin Drive, Baltimore, MD 21218, USA}

\author{John Chisholm}
\affiliation{Department of Astronomy, The University of Texas at Austin, 2515 Speedway, Stop C1400, Austin, TX 78712, USA}

\author[0000-0003-1427-2456]{Matteo Messa}
\affiliation{Observatoire de Genève, Université de Genève, Chemin Pegasi 51, Versoix CH-1290, Switzerland}
\affiliation{Department of Astronomy, Oskar Klein Centre, Stockholm University, AlbaNova University
Centre, SE-106 91 Stockholm, Sweden}

\author[0000-0002-0806-168X]{Linda J. Smith}
\affiliation{Space Telescope Science Institute, 3700 San Martin Drive, Baltimore, MD 21218, USA}

\author{Alessandra Aloisi}
\affiliation{Space Telescope Science Institute, 3700 San Martin Drive, Baltimore, MD 2121, USA}


\author[0000-0001-8289-3428]{Aida Wofford}
\affiliation{Instituto de Astronom\'{i}a, Universidad Nacional Aut\'{o}noma de M\'{e}xico, Unidad Acad\'{e}mica en Ensenada, Km 103 Carr. Tijuana-Ensenada, Ensenada 22860, M\'{e}xico}

\author[0000-0003-0724-4115]{Andrew J. Fox}
\affiliation{AURA for ESA, Space Telescope Science Institute, 3700 San Martin Dr, Baltimore, MD 21218, USA}

\author{Andrew Mizener}
\affiliation{Department of Astronomy, University of Massachusetts, 710 N. Pleasant Street, LGRT 619J, Amherst, MA 01002, USA}

\author{Christopher Usher}
\affiliation{Department of Astronomy, Oskar Klein Centre, Stockholm University, AlbaNova University
Centre, SE-106 91 Stockholm, Sweden}

\author[0000-0001-8068-0891]{Arjan Bik}
\affiliation{Department of Astronomy, Oskar Klein Centre, Stockholm University, AlbaNova University
Centre, SE-106 91 Stockholm, Sweden}

\author{Daniela Calzetti}
\affiliation{Department of Astronomy, University of Massachusetts, 710 N. Pleasant Street, LGRT 619J, Amherst, MA 01002, USA}

\author{Elena Sabbi}
\affiliation{Space Telescope Science Institute, 3700 San Martin Drive, Baltimore, MD 2121, USA}

\author{Eva Schinnerer}
\affiliation{Max Planck Institute for Astronomy, K̈onigstuhl 17,
69117 Heidelberg, Germany}

\author{Göran Östlin}
\affiliation{Department of Astronomy, Oskar Klein Centre, Stockholm University, AlbaNova University
Centre, SE-106 91 Stockholm, Sweden}

\author[0000-0002-3247-5321]{Kathryn Grasha}
\altaffiliation{ARC DECRA Fellow}
\affiliation{Research School of Astronomy and Astrophysics, Australian National University, Canberra, ACT 2611, Australia}  
\affiliation{ARC Centre of Excellence for All Sky Astrophysics in 3 Dimensions (ASTRO 3D), Australia}   

\author{Michele Cignoni}
\affiliation{Department of Physics - University of Pisa, Largo B. Pontecorvo 3, 56127, Pisa, Italy} 
\affiliation{INFN, Largo B. Pontecorvo 3, 56127, Pisa, Italy} 
\affiliation{INAF-Osservatorio Astronomico di Capodimonte, Via Moiariello 16, 80131, Napoli, Italy} 

\author{Michele Fumagalli}
\affiliation{Università degli Studi di Milano-Bicocca, Dip. di Fisica G. Occhialini, Piazza della Scienza 3, 20126 Milano, Italy}
\affiliation{INAF – Osservatorio Astronomico di Trieste, via G. B. Tiepolo 11, I-34143 Trieste, Italy}




\begin{abstract}


The CLusters in the Uv as EngineS (CLUES) survey is a Cosmic Origins Spectrograph (COS) campaign aimed at acquiring the 1130 to 1770 \AA\, restframe spectroscopy of very young ($<$20 Myr) and massive ($>10^4$ \msun) star clusters in galaxies that are part of the Hubble treasury program Legacy ExtraGalactic Uv Survey (LEGUS). 
In this first paper of a series, we describe the CLUES sample consisting of 20 young star clusters and report their physical properties as derived by both multi-wavelength photometry and far-UV (FUV) spectroscopy with Hubble Space Telescope (HST). Thanks to the synergy of the two different datasets we build a coherent picture of the diverse stellar populations found in each region (with sizes of 40 to 160 pc). We associate the FUV-brightest stellar population to the central targeted star cluster and the other modeled population to the diffuse stars that are included in the COS aperture. We observe better agreement between photometric and spectroscopic ages for star clusters younger than 5 Myr. For clusters older than 5 Myr, photometry and spectroscopy measurements deviate, with the latter producing older ages, due to the degeneracy of photometric models. FUV spectroscopy enables us to better constrain the stellar metallicities, a parameter that optical colors are insensitive to. Finally, the derived $E(B-V)$ are quite similar, with a tendency for FUV spectroscopy to favor solutions with higher extinctions. The recovered masses are in agreement within a factor of 2 for all the clusters.


\end{abstract}



\section{Introduction}
\label{sec:intro}




With the advent of the James Webb Space Telescope (JWST) and its unprecedented IR sensitivity, astronomers will soon be able to acquire integrated rest-frame far-ultraviolet (FUV) spectroscopy of high-redshift galaxies ($z>5$) for the first time. It is therefore fundamental to understand the stellar populations of FUV-bright star-forming galaxies in order to succeed with the next series of observational campaigns and maximise their scientific output. One of the most powerful and utilized instruments for this purpose is the Cosmic Origin Spectrograph \citep[COS][]{Green2012} onboard the Hubble Space Telescope (HST). FUV radiation unveils the recent star formation history of the galaxies as it is mainly produced by young massive stars ($M > 8$ \msun). Because massive stars produce feedback (stellar winds, radiation pressure, supernovae explosions and ionizing photons), they play an important role in galaxy evolution. Recent studies have measured the properties of massive stellar populations in samples of galaxies up to $z\sim2$ \citep[e.g.][]{chisholm19,Steidel16,reddy2020}, thanks to the employment of stellar population synthesis methods. These methods are able to model the FUV spectra to constrain the ages, metallicities, masses and reddening of the stellar populations.


In the nearby Universe, it is possible to observe the FUV spectra of galaxies with a higher signal-to-noise. This offers a unique window to study the properties of massive stars, the chemical evolution of galaxies, their feedback mechanisms and the production and escape of ionizing photons. The most recent and complete survey of this kind is CLASSY \citep[][, James et al. 2022]{Berg2022}, counting 45 nearby FUV-bright star-forming metal-poor galaxies.
In general, these types of data contain a wealth of information \citep[see e.g.][]{leitherer2020} that lies in four types of spectroscopic lines: (i) P-Cygni lines such as C IV 1548-1550Å, N V 1238-1242Å and Si IV 1393-1402Å that are produced in the stellar winds of massive stars and their strengths can vary as a function of metallicity and stellar age, offering an important diagnostic tool for determining the properties of the stellar populations; (ii) stellar photospheric absorption lines indicate the presence of older B stars ($> 7$ Myr) and are useful to constrain the star-formation history of the galaxy or star-forming region; (iii) absorption lines are produced also by the interstellar medium (ISM) along the line of sight to the FUV source (both the target galaxy ISM and the Milky Way ISM), which reveals the kinematics, the column density and the metallicity of the neutral and ionized gas; (iv) nebular emission lines can be equally important in determining the properties of stars and gas. 

Similar FUV studies to the ones mentioned above have been carried out also for star clusters and clustered star-forming knots within nearby galaxies. In general, FUV spectroscopy of young star clusters is a very powerful tool to derive their ages and masses \citep[e.g.][]{Wofford2011}. We expect young, massive, star clusters to dominate the light of high-redshift galaxies, whose starburst activity cannot be resolved. This represents an additional interest to investigate the properties of star clusters in order to understand galaxy evolution. The advantage of studying star clusters resides in simplifying the assumptions regarding the star formation histories and utilising them instead as proxies and tracers of star formation. FUV spectroscopy reveals the presence of very massive stars ($M > 100 $\msun) when dealing with unresolved stellar populations \citep{Wofford2014, Smith2016}. This can be used to derive information on their formation, stellar initial mass function of massive clusters, and the radiation ionizing the surrounding medium. 

The combination of high-spatial resolution HST imaging, optical and FUV spectroscopy has also helped to shed light on how well integrated FUV spectroscopy performs in reconstructing the recent star formation history and the feedback into the surrounding medium. \cite{Sirressi2022} report good agreement between the star cluster physical properties and the stellar feedback derived from HST multiband spectral distribution analysis (SED) and those recovered from FUV spectroscopy in two of the three star-forming knots analysed in the starburst galaxy Haro 11. Noticeable discrepancies between the two different methodologies are, however, found in the knot hosting the youngest star clusters with extinctions of E(B$-$V)$>$0.4 mag. Studies like \cite{Sirressi2022} are good examples of the power as well as of the limitations of studies focused only on FUV spectroscopic analyses.  This type of study is currently not possible for higher-redshift sources except for a few lensed galaxies \citep[e.g.][]{Citro2021,Vanzella2019}, where the magnification of the gravitational lens makes observable dense star-forming regions (sometimes called clumps) in the early Universe. 

FUV spectroscopy of young stellar clusters has also proven a powerful tool for measuring the metallicity of stellar populations. Comparisons among metallicity gradients derived with HII region abundance analyses, analysis of red supergiant cluster spectroscopy, and FUV spectral fitting of young star clusters show very good agreements, except for the central region of the target galaxy, suggesting that these different barionic tracers can aid to the understanding of metal enrichment of galaxies \citep[][]{hernandez2019}. The agreement between the metallicities derived from the young stellar populations (in FUV) and the ionized nebular phases (from optical abundance line tracers) points toward a timescale for mixing the newly synthesized metals longer than the lifetimes of massive stars, which are considered the primary polluters in star-forming regions \citep{Hernandez2021}. 




Here we present the CLusters in the Uv as EngineS (CLUES) sample, which comprises FUV COS spectroscopy of 20 young star clusters hosted in 11 nearby star-forming galaxies. The scientific goals of this HST program will be presented in a series of papers. \emph{(i)}. In this work, we study the properties of the stellar populations (ages, metallicities, internal reddening to the clusters and masses) using a variety of methodologies such as photometric modelling, single and double population spectroscopic modelling as well as multiple population spectroscopic modelling. \emph{(ii)}. In a follow-up analysis, (Mizener et al. in prep.) we complement the analysis of the stellar populations by fitting the LEGUS HST multiband photometry and testing various models including the Binary Population and Spectral Synthesis (BPASS). \emph{(iii)}. In Sirressi et al. (in prep.) we study the dynamical properties of the intervening neutral gas along the line of sight to each cluster, most importantly the gas kinematics and how it correlates with the stellar feedback from the clusters. \emph{(iv)}. Neutral ISM abundance analysis will be presented in a subsequent paper \emph{(v)}. Additionally we will analyse the physical properties of the Milky Way ISM absorption lines.

The paper is organised as follows. We describe the sample and how it was selected in Section \ref{sec:select}. In Section \ref{sec:COSdata} we present the available COS data and the data reduction, whereas in Section \ref{sec:phot_data} we present the HST photometry that we use as complementary data for our analysis. We report the analysis of the FUV spectroscopy in Section \ref{sec:fit} and the results of the whole analysis in Section \ref{sec:results}. Discussions and conclusions are drawn in Section \ref{sec:discuss} and \ref{sec:conclusions}.

\section{Sample description \& selection}
\label{sec:select}
The CLUES sample is drawn from the large parent sample of star clusters detected by the Legacy ExtraGalactic UV Survey \citep[LEGUS,][]{calzetti2015}. The galaxies from which the CLUES targets were selected, located at a distance between 3 and 13 Mpc cover a wide range of star formation properties (SFR = 0.1 - 6.8 $\msunyr$), metallicity (12 + log(O/H) = 8.0 - 9.0) and morphology (grand-design arms, circumnuclear starburst rings, flocculent spirals, dwarf starburst and tidal features) representative of the LEGUS parent sample \citep{calzetti2015}. 

\begin{figure*}
    \centering
    \includegraphics[width=0.45\textwidth]{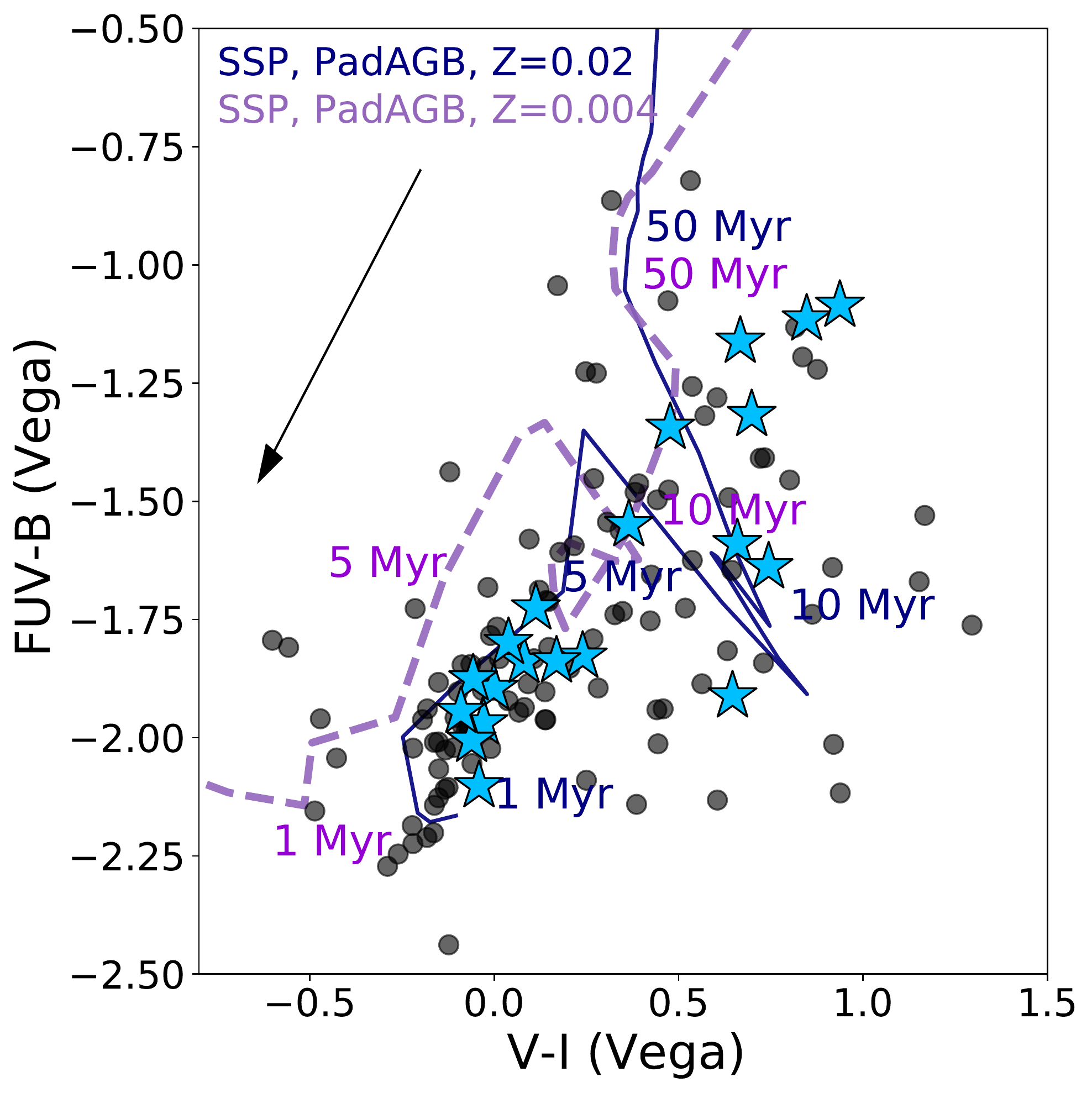}
    \includegraphics[width=0.45\textwidth]{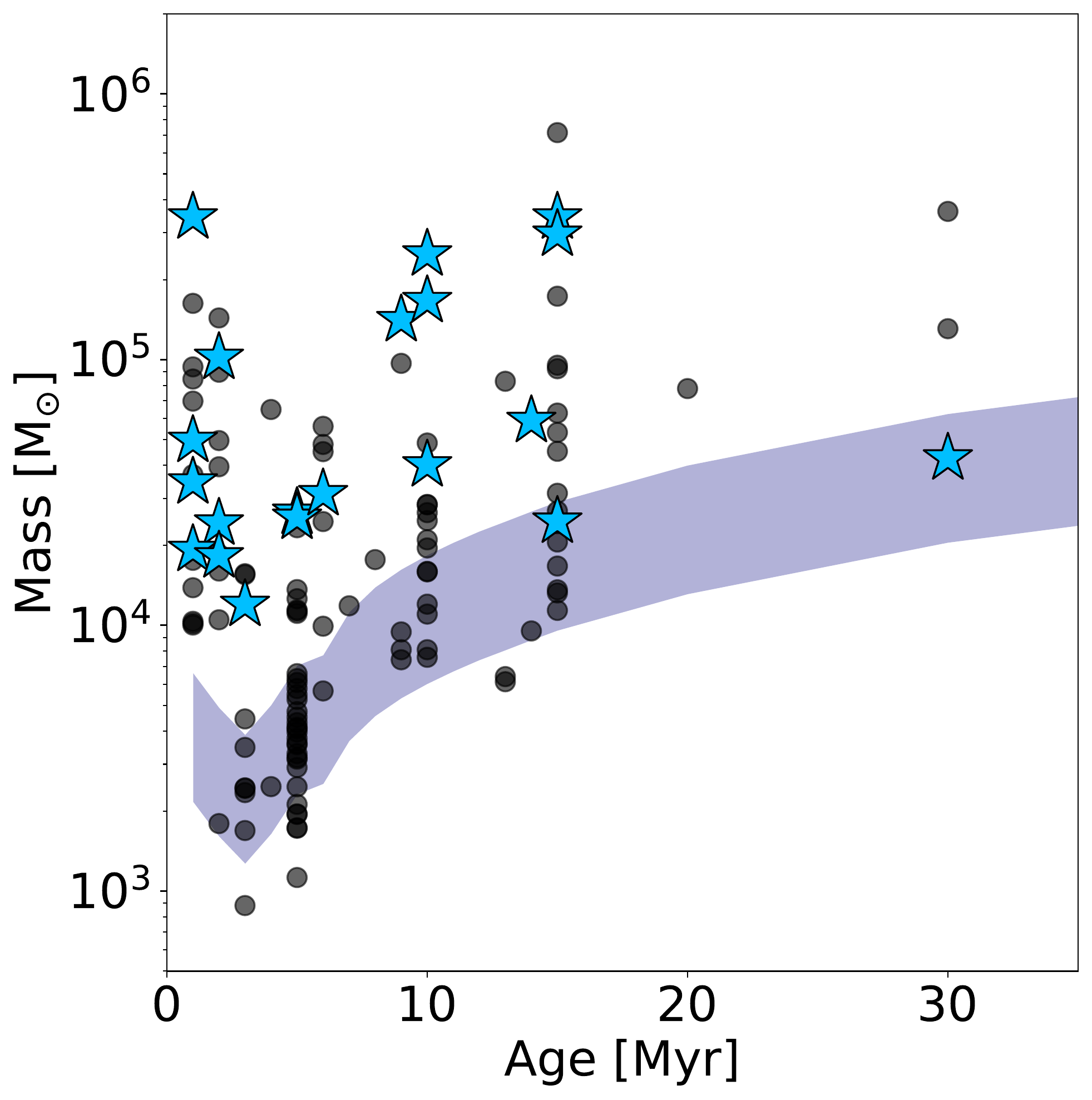} 
    \caption{Parameter space spanned by the clusters belonging to the CLUES sample (cyan stars). The clusters that pass the first selection (class 1 and 2 with m$_{F275W} <$ 18 mag, $E(B-V)\leq$ 0.3 mag, age $\leq$ 30 Myr) are shown in both panels as grey dots. The left panel shows the FUV color, F275W$-$F438W (or F435W) vs. the optical color, F555W (or F606W)$-$F814W, of the clusters. Yggdrasil evolutionary tracks spanning the metallicity range of the LEGUS sample are included and the main age steps outlined. The arrow shows in which direction clusters would move if corrected by an internal reddening corresponding to $E(B-V)=0.3$ mag. In the right panel, we show the age vs. mass diagram of the sample. The violet band shows the detection limit corresponding m$_{F275W}=$18 mag as function of the metallicity (Z$=0.004$ and 0.02) and distance range of the LEGUS galaxies.}
    \label{fig:selection}
\end{figure*}

\begin{figure*}
    \centering
    \includegraphics[width=\textwidth]{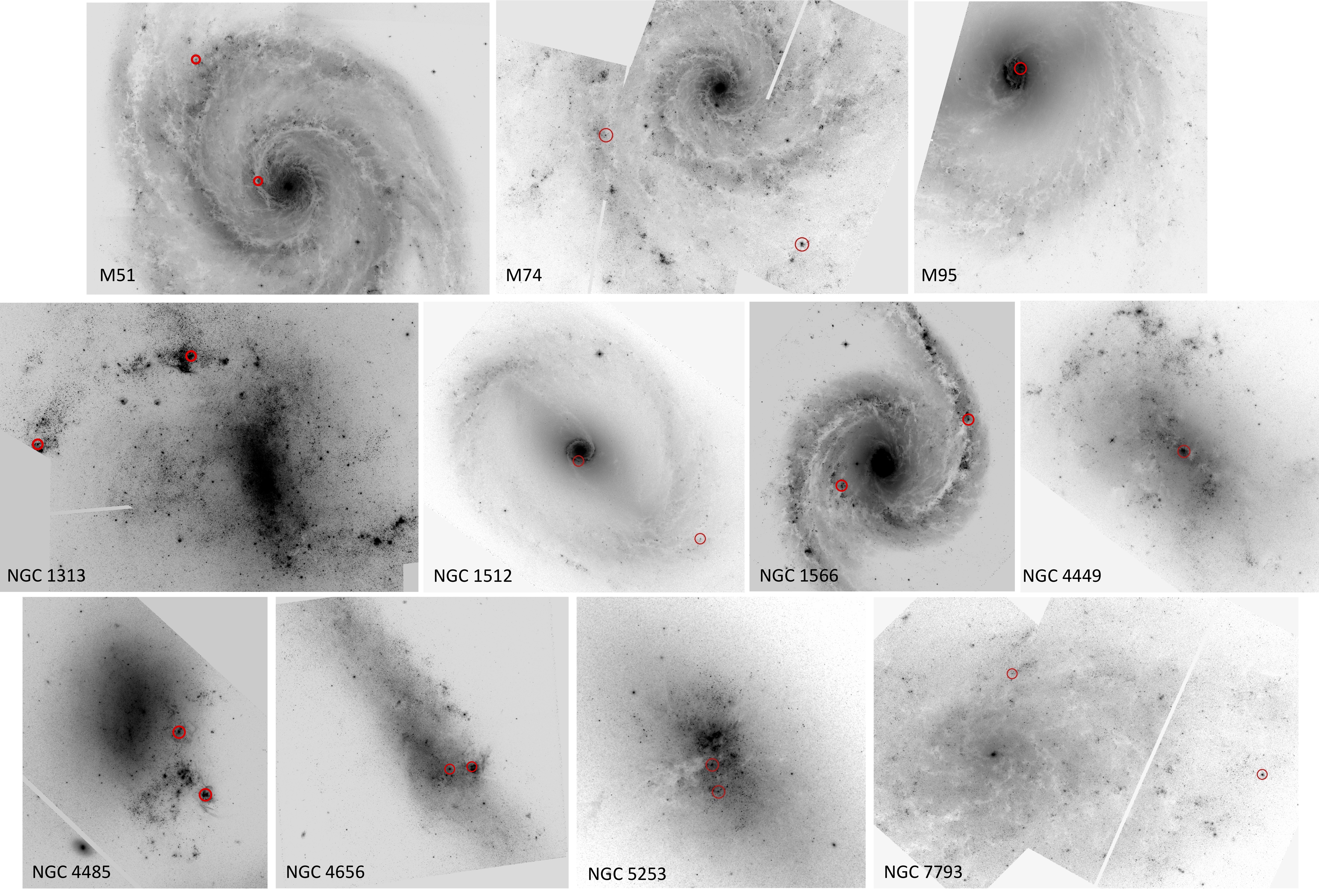}
     \caption{Overview of the CLUES sample. On top of the LEGUS F555W (F435W for M74) images of the targets we mark with red circles the positions of the regions where FUV spectroscopy with COS has been acquired and presented in this work. Images are all rotated north-up and intensities are in logarithmic scales. See Table~\ref{tab:sample} for a summary of the galactic properties.}
    \label{fig:sample}
\end{figure*}

The star cluster catalogues of 37 star-forming galaxies (consisting of dwarfs, spirals and interacting systems) were inspected to select star cluster candidates with apparent magnitude in the NUV filter, F275W, brighter than 18.0 mag (before Galactic extinction correction). 
The first selection included clusters that have been classified as compact during the visual inspection \citep[class 1 and 2, see][]{Adamo2017}, that have internal reddening $E(B-V)\leq$0.3 mag, and ages younger than 30 Myr from SED fitting \citep[photometric age estimates have average errors of 0.2 dex,][]{Adamo2017}. The flux density cut at 18 mag in the F275W corresponds to a mass cut of about $\sim 10^4 \msun$ at these age ranges (see Figure~\ref{fig:selection}). This restricted our selection to the massive, and thus more effective, sources of feedback in each galaxy. This first selection resulted in 125 young star clusters (YSCs), which were then visually vetted for contaminating clusters (class 1 and 2) located within the aperture of the COS spectrograph. We also removed clusters located in galaxies that have Galactic foreground extinction larger than 0.1 mag (see Table~\ref{tab:sample}). When possible, we prioritized the selection of galaxies with two clusters that satisfy all the conditions described above but located at different distance from their center. This was desired in order to study potential systematic differences in cluster properties as a function of location of the clusters in the host galaxy. The final CLUES sample includes 20 YSCs (distributed in 11 galaxies, see Table~\ref{tab:sample} and Figures~\ref{fig:selection} and \ref{fig:sample}) with a photometric age $\leq$ 30 Myr, mass $\geq 10^4\,\msun$ (from LEGUS analysis) and dominating the light emission within a radius of 0.2" from the center of the COS aperture (hence the contribution to the COS spectrum of each star cluster is significant). In Figure~\ref{fig:selection}, we show that the selection of the CLUES targets is not biased against specific colors, ages or masses, but it is representative of the parameter space occupied by the 125 cluster candidates (underlying grey dots) that passed the first selection stage.
Three of these clusters already have archival COS data available (NGC-4449-YSC1, NGC-5253-YSC1, NGC-5253-YSC2). All the key age phases are sampled: 1-3 Myr, dominated by the feedback of massive stars (photo-ionization and wind); 
4-5 Myr, dominated by the feedback from Wolf-Rayet stars;
6-20 Myr, when most of the mechanical energy is produced by type II, core-collapse SNe \citep[see][for a review]{krumholz2014}. The stellar age estimates above refer to single stellar populations but will need to be adjusted once binaries are included as most massive stars are now well established to be in close binaries \citep{Mason2009,Dunstall2015}.

Table \ref{tab:sample} lists the names of the sources of the CLUES sample, the properties of the host galaxies, as well as the observation set up and S/N reached in each spectrum. Figure~\ref{fig:sample} shows the location of the targets within their respective galaxies, spanning a range of different morphologies.  
All LEGUS science frames and cluster catalogues are publicly available at https://legus.stsci.edu/index.html.

\begin{table*}
\centering
\begin{tabular}{lcccccccccc}
\hline
 source name   &        $z$ &   $v$  &   $D$  &   $SFR$  &   12+log(O/H) &   $E(B-V)_{MW}$ & G130M & G160M & m$_{F275W}$\\

 &          &    [km/s] &    [Mpc] &   [$\msunyr$] &  &   [mag] & [sec] (S/N) & [sec] (S/N) & [mag]\\
 (1) & (2) & (3) & (4) & (5) & (6) & (7) & (8) & (9) & (10)\\
\hline
 M51-1         & 0.00148 &        444 &      7.66 &          6.88 &           8.8 &       0.031 & 2275 (22) & 2601 (11) & 15.3\\
 M51-2         & 0.00110 &        330 &      7.66 &          6.88 &           8.8 &       0.031 & 5212 (21) & 8439 (12) & 17.1\\
 M74-1         & 0.00226 &        677 &      9.90 &          3.67 &           8.7 &       0.062 & 5048 (19) & 8050 (11) & 17.5\\
 M74-2         & 0.00223 &        669 &      9.90 &          3.67 &           8.7 &       0.062 & 5048 (13) & 8050 (7) & 17.5\\
 M95-1         & 0.00260 &        779 &     10.00 &          1.57 &           9.2 &       0.025 & 2228 (23) & 2465 (12) & 16.2\\
 NGC1313-1     & 0.00164 &        492 &      4.39 &          1.15 &           8.4 &       0.096 & 1751 (27) & 3505 (17) & 15.0\\
 NGC1313-2     & 0.00161 &        484 &      4.39 &          1.15 &           8.4 &       0.096 & 5561 (20) & 9039 (13) & 16.4\\
 NGC1512-1     & 0.00314 &        942 &     11.60 &          1.00 &           8.8 &       0.009 & 2397 (9) & 2765 (3) & 17.7\\
 NGC1512-2     & 0.00271 &        812 &     11.60 &          1.00 &           8.8 &       0.009  & 5113 (23) & 8306 (13) & 17.1\\
 NGC1566-1     & 0.00516 &       1547 &     13.20 &          5.67 &           9.1 &       0.008 & 2429 (15) & 2860 (11) & 17.8\\
 NGC1566-2     & 0.00505 &       1515 &     13.20 &          5.67 &           9.1 &       0.008 & 4937 (29) & 8586 (16) & 17.0\\
 (*)NGC4449-1     & 0.00084 &        253 &      4.31 &          0.94 &           8.3 &       0.017 & 1736 (45) & 944 (17) & 15.5\\
 NGC4485-1     & 0.00147 &        440 &      7.60 &          0.25 &         -   &       0.019 & 2463 (26) & 5529 (17) & 16.4\\
 NGC4485-2     & 0.00151 &        453 &      7.60 &          0.25 &         -   &       0.019 & 2434 (21) & 5529 (14) & 16.6\\
 NGC4656-1     & 0.00214 &        642 &      5.50 &          0.50 &           8.1 &       0.012 & 2845 (20) & 4528 (12) & 16.3\\
 NGC4656-2     & 0.00215 &        643 &      5.50 &          0.50 &           8.1 &       0.012 & 980 (26) & 1063 (14) & 14.4\\
 (*)NGC5253-1     & 0.00141 &        422 &      3.15 &          0.10 &           8.2 &       0.049 & 4694 (41) & 3473 (21) & 15.2\\
 (*)NGC5253-2     & 0.00145 &        433 &      3.15 &          0.10 &           8.2 &       0.049 & 3013 (50) & 6296 (25) & 16.1\\
 NGC7793-1     & 0.00077 &        230 &      3.44 &          0.52 &           8.6 &       0.017 & 1540 (24) & 3042 (15) & 15.4\\
 NGC7793-2     & 0.00101 &        302 &      3.44 &          0.52 &           8.6 &       0.017 & 3525 (26) & 6705 (16) & 17.1\\
\hline
\end{tabular}
\caption{Source names of the clusters in the CLUES sample and properties of their host galaxies. Column (1): source name. Columns (2) and (3): redshift and systemic velocity of the cluster as derived from FUV photospheric absoprtion lines (see Section \ref{sec:fit}), Column (4): luminosity distance of the host galaxy from \cite{calzetti2015}. Column (5): star formation rate of the host galaxy \citep[from][]{calzetti2015}. Column (6): oxygen abundances of the host galaxy \citep[from][]{calzetti2015}, mean value when multiple measurement are available. Column (7): reddening attenuation due to the Milky Way dust along the line of sight to the host galaxy, taken from the NASA/IPAC Extragalactic Database (NED), \cite{schlafly2011}}. Columns (8) and (9): exposure time of observations with gratings G130M and G160M and (in parenthesis) signal-to-noise of the two rebinned (0.4 Å) spectra calculated with a median over the portions of the spectrum included in our fit (see Section \ref{sec:fit}). Column (10): magnitude of the target cluster in the F275W Hubble filter as reported in the LEGUS catalogue. The targets with an asterisk (*) in front of their name, have been observed using different position angles.
\label{tab:sample}
\end{table*}

\begin{figure*}
    \centering
    \includegraphics[width=\textwidth]{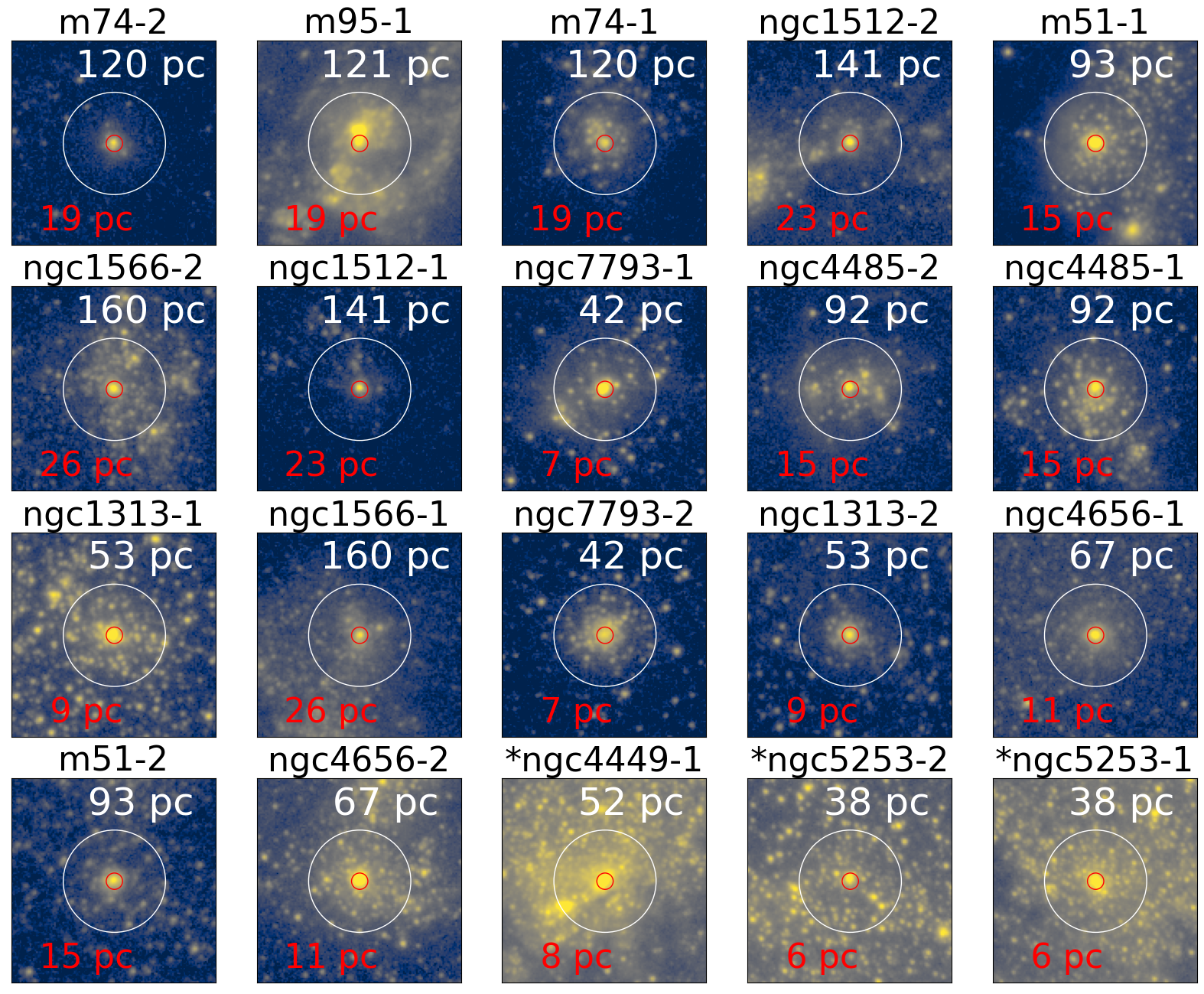}
     \caption{F275W images of the CLUES targets sorted by increasing cluster age as derived by FUV spectroscopy. The larger white circle represents the COS aperture of 2.5\arcsec diameter, which corresponds to the physical size printed with the same color in each frame. The smaller red circle has a diameter of 0.4\arcsec (physical size printed in red) within which lies the targeted young star cluster. The number at the end of the target name is the identifier of the cluster as for most of the host galaxies we have selected two young star clusters as part of the CLUES sample. The targets with an asterisk (*) in front of their name, have been observed using different position angles.}
    \label{fig:images}
\end{figure*}

Figure \ref{fig:images} illustrates a close zoom-in in the F275W HST images of the targets. The circular COS aperture (larger white circle, corresponding to 2.5\arcsec diameter) encloses the cluster light at the center (within the red circle of 0.4\arcsec diameter) and in most of the targets an elevated clustering of stars, as expected because of the young ages of the star-forming regions. The diameter size of such aperture in physical units is printed in each frame of the figure and depends on the distance of each target. We are targeting regions that have sizes between 40 (for the closest target at $\sim$4 Mpc, NGC-5253) and 160 pc (the farthest target at 13 Mpc, NGC-1566). For a given aperture, the physical size depends on the distance of the target, which was reported in \cite{calzetti2015}.

\section{COS data reduction}
\label{sec:COSdata}

We summarize here the data acquisition and the reduction steps that produce the final combined spectrum of each target.

\subsection{Grating setup and observing strategy}

The data presented in this paper were obtained with the COS instrument onboard HST \citep[][]{Green2012} as part of the program ID 15627 (PI: Adamo). Three sources, namely NGC-4449-YSC1, NGC-5253-YSC1 and NGC-5253-YSC2, were observed with the same instrument as part of an earlier program (ID 11579, PI: Aloisi). All the {\it HST/COS} data used in this paper can be found in MAST: \dataset[10.17909/n9p3-n688]{http://dx.doi.org/10.17909/n9p3-n688}. Since the positions of the clusters are known to a precision higher than 0.4\arcsec, we used the ACQ/IMAGE setting to perform the target acquisition. Target imaging acquisition in the NUV was done with MIRROR/A in 11 of the 17 targets (those fainter than 17 mag), with the remaining targets using MIRROR/B. Spectra were acquired with the lifetime position 4 of the detector. Each cluster is observed using two or three different grating configurations (G130M-1291, G160M-1589, G160M-1600) ensuring that all the spectral lines of interest would not fall into the gap of each CENWAVE. Spectroscopy in the two gratings (G130M and G160M) is performed during the same visit for each target. Single spectra in the G130M setting were acquired at FP-POS $=$ 3 and 4, while the G160M single spectra were acquired with FP-POS 1, 2, 3, and 4. The two combined settings provide an overall spectral coverage between 1130 Å and 1770 Å (rest-frame) for all sources.

\subsection{Extraction, spectral calibration, final combined spectra}

The observations were retrieved from the Mikulski Archive for Space Telescopes (MAST) and calibrated with the HST pipeline, CALCOS V3.3.10 \citep{Johnson2021}. For those datasets where more than one association file is collected per configuration, or where multiple central wavelengths are observed for a single target and grating, we combine the individual x1d files using the IDL code by \cite{Danforth2010}. The software allows us to combine spectra from multiple exposures weighted by exposure times and interpolate onto a common wavelength vector. 


The flux errors returned by the pipeline are overestimated, especially for low S/N sources \citep[][]{Henry2015}. The pipeline error vector has the correct structure, however its level needs to adjusted, as it is not representative of the dispersion in the data. We therefore measure the errors by calculating the standard deviation in a window of 50 spectral pixels around each wavelength and fitting a high order polynomial (degree=7) to the vector of standard deviations. We then rescale the actual error vector from the pipeline to the appropriate level. In this computation we mask out absorption lines and bright geocoronal lines. We estimate errors up to a factor of 2 smaller than the CALCOS pipeline errors.


For each source, the spectra of the two different gratings G130M and G160M are combined into a single spectrum. To do this, we first define a common wavelength grid of bin size 0.4 Å (see Section \ref{sec:res}) and then we resample the individual spectra onto the common grid. As a next step we simply concatenate resampled spectra and use a weighted-average for the overlapping region between $\sim 1400-1420$ Å (observed frame). The weighted average is also used for the two G160M gratings of different CENWAVE (G160M-1589 and G160M-1600), for those sources who have both observation settings. 
The average continuum S/N of the rebinned FUV spectra for the two gratings (G130M and G160M) are listed in Table \ref{tab:sample}. The lowest value S/N$\sim$3 is for gratings G160M of the target NGC-1512-YSC1, however we do not exclude this source from our spectroscopic fits as the P-Cygni lines are prominent in its spectrum and the fit converges.

One caveat regarding the archival data is that 3 of our targets (NGC-5253-YSC1, NGC-5253-YSC2, NGC-4449-YSC1) are observed with the two gratings G130M and G160M at different times, using different position angles. This means that the two gratings have different contribution from the diffuse stellar population. For this reason, the quality of the spectral fit for these targets is inferior (see Figure \ref{fig:app3} and \ref{fig:app4} of the appendix), especially at the longest wavelengths.\\

\subsection{Effective spectral resolution}
\label{sec:res}
Star clusters at the distances of the LEGUS galaxies have full-width half maximum (FWHM) larger than a stellar point spread function. Thus we expect the resulting effective spectral resolution of the COS observations to be lower than the one defined by the COS Line Spread Function (LSF). In order to estimate the effective spectral resolution we use the NUV acquisition images in the following way. We measure the profile of the NUV acquisition image along the dispersion direction and re-scale it to match the FUV pixel-size. We fit this profile with a Lorentzian function and perform the convolution of the tabulated COS LSF with the fitted acquisition image profile. Finally we measure the FWHM of the convolution as an estimate of the effective spectral resolution. In this work, we fit the FUV spectra with single stellar population models in order to determine cluster physical properties. As detailed in Section~\ref{sec:fit}, we use STARBURST99 evolutionary tracks \citep{leitherer1999, leitherer2014}, which have a spectral resolution of 0.4 \AA. The measured FWHM of all the CLUES targets have spectral resolution better than 0.4 \AA. The spectral effective resolutions will be presented and discussed in a forthcoming work (Sirressi et al. in prep.) focused on the kinematics of the intervening ISM absorption lines. For this reason, in this work we decide to re-sample the observed FUV spectra to 0.4 Å and match the spectral resolution of the modeled spectra as described in Section \ref{sec:FUV_spec}.

\subsection[Fit of the HI absorption]{Fit of the Ly-$\alpha$ absorption wings}
\label{sec:lya}
As a last step, before the spectral analysis, we fit the Lyman-$\alpha$ (Ly-$\alpha$) damped wings belonging to both the science target and the foreground Milky Way. This step is necessary because the red wings of the Ly-$\alpha$ absorption affects the shape of the N V 1238-1242Å P-Cygni doublet, one of the age sensitive lines in the spectra. 

The Ly-$\alpha$ profile fitting analysis includes the normalization of the individual COS observations. The continuum normalization is done by interpolating between
nodes chosen to avoid any absorption features present (stellar or ISM). We apply a cubic spline interpolation, adjusting the tension of the spline curve as needed. The nodes positions was based on STARBURST99 \citep[SB99, ][]{leitherer1999,leitherer2014} models of instantaneous bursts with age, metallicity and reddening attenuation taken from the LEGUS analysis \citep[][]{calzetti2015}. \par
The column densities for the Ly-$\alpha$ absorption line at $\lambda$ = 1215.671 Å are derived by fitting Voigt profiles using the Python tool VoigtFit \citep[][]{Krogager2018}. This software allows us to account for the instrumental line-spread function (LSF), as well as the broadening introduced by the extension of the source within the 2.5" COS aperture. Following an approach similar to \cite{hernandez2020}, the intrinsic COS LSF profiles are broadened to accurately account for the source extension adopting the final spectral resolution obtained using their equation (1). \par
Given the nature of the objects analyzed here, specifically their close proximity, the Ly-$\alpha$  absorption originating from the Milky Way (MW) ISM is heavily blended with that from the targets themselves. Similar to the approach adopted in \cite{Hernandez2021} we simultaneously fit the extragalactic and Galactic Ly-$\alpha$  profiles. The red wing of the broad Ly-$\alpha$  profile is used to constrain the HI column density fit of the targets, and the blue wing is used to constrain the MW HI column density. These measurements will be reported in upcoming papers (Sirressi et al in prep.). \par

\section{Ancillary data \& photometric analysis}
\label{sec:phot_data}
All the targets have the standard LEGUS HST imaging coverage consisting of 5 broadband filters: WFC3/F275W (NUV) and F336W (U), ACS or WFC3 in F435W or F438W (B), F555W or F606W (V), F814W (I),  sampling the cluster spectral energy distribution (SED) from the NUV to the NIR \citep[see as reference][]{calzetti2015}. Data are drizzled to a pixel scale of 0.04\arcsec and aligned with respect to the B band. Aperture photometry (with radius of 4 to 6 pixels depending on the distance of the galaxy) are performed to determine the fluxes of the clusters in each band. The final cluster photometry is corrected for Milky Way foreground reddening and loss due to finite aperture \citep[see][as reference]{Adamo2017}. The SED of each cluster is fitted with a Yggdrasil single stellar population model \citep{zackrisson2011}, which account also for nebular emission lines and continuum. As presented in \citet{Adamo2017}, several different output catalogues are produced, assuming different stellar libraries, extinction laws, and metallicities. We stress that, while the selection of the CLUES targets is made using the Padova evolutionary tracks and assuming a Cardelli extinction law (see Section~\ref{sec:select}), in the analysis presented below, we use the best fitted cluster parameters obtained with similar assumptions as those made to fit the FUV spectra. We extract the best photometric SED cluster parameters obtained by fitting cluster fluxes estimated with average aperture corrections, using Geneva non-rotating evolutionary tracks from STARBURST99,  assuming a fully populated \cite{kroupa2001} initial mass function (IMF) and differential starburst reddening \citep{calzetti2000}. The best-fit parameters extracted from the LEGUS catalogues are listed in Table~\ref{tab:fit} for each cluster. We note that the metallicity is not a free parameter in the fit of the cluster photometric SEDs but it is fixed to a value that is closest to the average value of the galaxy (see Tables~\ref{tab:sample} and \ref{tab:fit}).

\section{Spectral analysis}
\label{sec:fit}

\label{sec:FUV_spec}

To determine the stellar population physical parameters of the regions targeted with COS, we fit the FUV spectroscopy using two different methods. In the first method, we assume that two single stellar populations (the cluster and the surrounding diffuse stellar population) contribute to the FUV flux of the targeted region.  
In the second method, we fit two or more single stellar populations until a convergence is reached in the fit. The first method is described in section 5.1, while the second method is described in section 5.2. As we will further discuss in Section \ref{sec:phot}, the spectroscopy of our target star clusters is contaminated by the stellar emission from a diffuse stellar population included in the COS aperture. This supports the usage of the second method with more than one or two stellar populations, although it introduces a higher number of fitting parameters and therefore degeneracies between them. We pre-process our spectra for the fitting process as follows. 
In the first step, the spectra are re-binned to 0.4 Å to match the resolution of the SB99 models \citep[SB99, ][]{leitherer1999,leitherer2014}.
Secondly, the photospheric FUV lines C III 1176Å and C III 1247Å are used to measure the FWHM (to account for the broadening caused by stellar kinematics) and the redshift of the target (which due to galactic rotation differs from the redshift of the galaxy, see Table~\ref{tab:sample}). Before fitting each target, the spectral models are redshifted by the corresponding measured redshift and the observed spectra are corrected for the reddening attenuation of the Milky Way (produced from NED and tabulated in Table~\ref{tab:sample}) using the CCM89 law \citep[][]{Cardelli1989}. 

All the FUV spectra of our sample feature a broad Ly-$\alpha$ absorption around 1216 Å that can contaminate the N V 1238-1242Å P-Cygni doublet (wavelengths given in the rest-frame). After fitting the Voigt profile of the Ly-$\alpha$ as described in Section \ref{sec:lya}, we divide the normalised science spectrum by the normalised Voigt model. The resulting spectrum is then rescaled by using the same normalisation constant, obtaining in this way a rectified spectrum which overlaps with the initial spectrum except in the region around the Ly-$\alpha$ emission.

\begin{figure*}
    \centering
    \includegraphics[width=\textwidth]{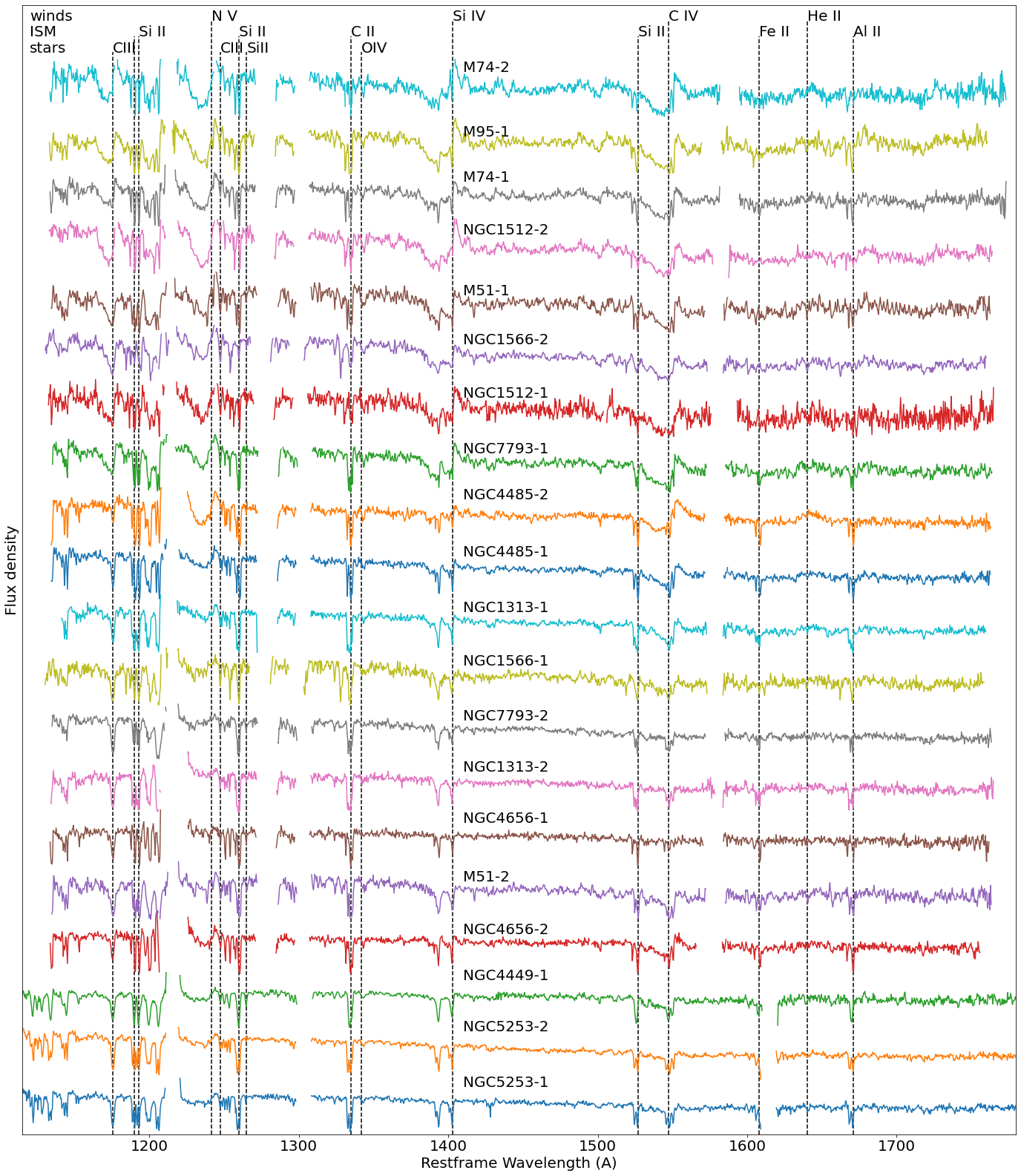}
     \caption{Atlas of the CLUES FUV spectra at 0.4 \AA\, spectral resolution, sorted by age (see text). The vertical dashed lines indicate some of the photospheric lines, stellar wind P-Cygni lines and absorption lines due to the interstellar medium along the line of sight. These marked lines are labelled on top of the figure. The regions of the spectra that are not included in our analysis such as the Ly-a emission, the O I geocoronal line and the detector gaps at about 1300 Å and 1600 Å are omitted for clarity. See Section \ref{sec:FUV_spec} for the full description of the mask used when modelling the FUV spectroscopy. }
    \label{fig:atlas}
\end{figure*}

Figure \ref{fig:atlas} illustrates the atlas of the CLUES spectra in the restframe wavelengths, corrected for the Ly-$\alpha$ damping winds, and re-binned to 0.4 \AA. The spectra in the figure are sorted by increasing age of the FUV-bright stellar population (\textit{Pop 1}, see Sec \ref{sec:double}) of the spectroscopy model. The Ly-$\alpha$ and geocoronal OI emission as well as the detector gaps have been masked for visualisation purposes. The reader can appreciate the richness of information available in the spectra. One can easily see photospheric and broad stellar wind P-Cygni lines typical of very young stellar populations. Absorption lines due to the intervening interstellar medium (ISM) along the line of sight are also visible and will be analysed in a forthcoming paper. 

Finally, we visually inspect each FUV spectrum of the CLUES sample in order to define a target-specific mask that excludes certain spectral regions when running the fit. The masked regions include: Ly-$\alpha$ emission, O I geocoronal line, detector's gaps at $\sim$ 1300 Å and $\sim$ 1600 Å (observed frame), flagged spectral pixels (with insufficient data quality), and regions with ISM absorption lines (as our models implement only the stellar physics). The applied masks are displayed in Figures \ref{fig:example} and \ref{fig:app} in the Appendix.

\subsection{Single and double stellar population}
\label{sec:double}
The properties of the stellar populations are inferred using SB99 theoretical stellar libraries \citep[SB99, ][]{leitherer1999,leitherer2014} and a code that constructs a stellar spectrum as a continuous function of age, metallicity, mass and reddening attenuation. A version of the code similar to the one used in this work is described in \cite{Sirressi2022}. We model the FUV spectra with stellar libraries using the Geneva stellar evolutionary tracks with high mass-loss rates \citep{Meynet1994}, Salpeter IMF\footnote{The values of the masses derived from spectroscopy have been multiplied by 0.68 according to the conversion from a Salpeter IMF to a Kroupa IMF \citep[][]{kennicutt2012}, to make the comparison consistent with the photometry values.} with cut-offs at 0.1 and 120 $\rm M_{\odot}$, non-rotating stars and with a single-burst star-formation history. These models include both the continuum emission produced by stars and the continuum emission produced by the nebular gas. To derive the extinction we use the starburst attenuation law \citep{calzetti2000}.

We fit the FUV spectrum of each cluster with a model assuming that there are two stellar populations. Each modeled population has 4 fitting parameters: stellar age, metallicity $Z$, mass and reddening attenuation $E(B-V)$. The priors on the parameters are identical for the two stellar populations and are set up consistently with the ranges of the values derived from photometry. Ages can vary between 1 and 50 Myr, metallicity between 0.004 and 0.04 if the host is a spiral galaxy, otherwise between 0.001 and 0.008 if it is a dwarf galaxy, maximum mass is set to $10^7\,\msun$ (considering we are fitting regions which are between tens and a few hundreds of parsec across), reddening attenuation E(B$-$V) between 0.01 and 0.8 mag. Stellar libraries are produced following a grid of models for the metallicities, with values of 0.001, 0.004, 0.008, 0.020, 0.040, with an age step of 0.1 Myr. Before the minimisation, the models are spline-interpolated such that a synthetic spectrum can be found for any given value of the parameters within their priors and not only for the value of the grid. The mass of the stellar population is given by the normalisation constant needed to match the model spectrum (corresponding to $10^6\,\rm\msun$) to the observed spectrum. 

Before fitting each target, the spectral models are convolved with a Gaussian shape with the same FWHM as measured in the observed spectra, using either C III 1247 or C III 1176. We define \textit{Pop 1} as the model component with a higher flux contribution at 1276 Å, and \textit{Pop 2} as the model with the lower flux. We motivate the inclusion of a second population to account for the fact that the stellar population surrounding the star cluster might not have the same physical parameters. A complete description of this software will be provided in Hayes et al. (in prep.). 
We further analyse the relative contribution of the cluster light with respect to the surrounding stellar population located within the COS aperture in Section \ref{sec:phot}.
In order to assert whether the two-population model gives a better fit to the spectra of the targets than a single population model, we use the Akaike information criterion (AIC). We compare the AIC estimators of the fit with a single population and the fit with two populations. The AIC estimator, founded on information theory, allows us to estimate the relative information loss when only a single population model is used rather than a double population model. For a formal definition of the estimator and its applications in astrophysical models see \cite{Liddle2007}. We find that the two-population models give a higher quality description of the data, except in four cases: M-74-YSC2, NGC-7793-YSC1, NGC-1566-YSC1, NGC-1566-YSC2. For these sources we quote only the values of the single-population fit. In 16 out of 20 cases, a second population is required to accurately reproduce the observed level of FUV continuum and the detected P-Cygni lines simultaneously. Table \ref{tab:aic} lists the AIC numbers for the models with a single stellar population compared to a double stellar population.

\begin{table}
\centering
\begin{tabular}{lrr}
\hline
 source name   & AIC (single pop.) &   AIC (double pop.)  \\
\hline
 \textbf{M74-2}     & \textbf{1110} & \textbf{1110} \\
 M95-1     & 2687 & 2653 \\
 M74-1     & 1361 & 1323 \\
 NGC1512-2 & 2490 & 2474 \\
 M51-1     & 2507 & 2490 \\
 \textbf{NGC1566-2} & \textbf{2418} & \textbf{2418} \\
 NGC1512-1 &  854 &  595 \\
 \textbf{NGC7793-1} & \textbf{2277} & \textbf{2277} \\
 NGC4485-2 & 1445 & 1196 \\
 NGC4485-1 & 1382 & 1341 \\
 NGC1313-1 & 1632 & 1629 \\
 \textbf{NGC1566-1} & \textbf{1425} & \textbf{1703} \\
 NGC7793-2 & 2103 & 1689 \\
 NGC1313-2 & 1241 & 1086 \\
 NGC4656-1 &  950 &  942 \\
 M51-2     & 1943 & 1895 \\
 NGC4656-2 & 1557 & 1423 \\
 NGC4449-1 & 1855 & 1629 \\
 NGC5253-2 & 1865 & 1731 \\
 NGC5253-1 & 1703 & 1459 \\
\hline
\end{tabular}
\caption{AIC values of single and double stellar population models for each target. The targets marked in bold are those where the double population model does not provide an improvement relative to the single population model. The definition of the AIC estimator can be found in \cite{Liddle2007}.}
\label{tab:aic}
\end{table}

\begin{figure*}
    \centering
	\includegraphics[width=0.40\textwidth]{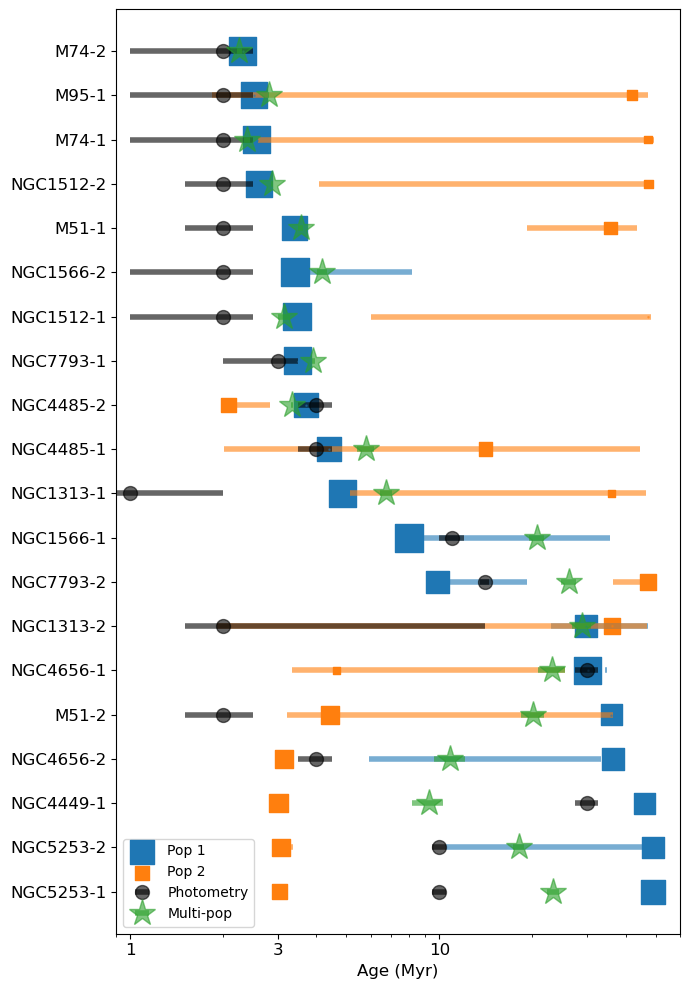}
    \includegraphics[width=0.40\textwidth]{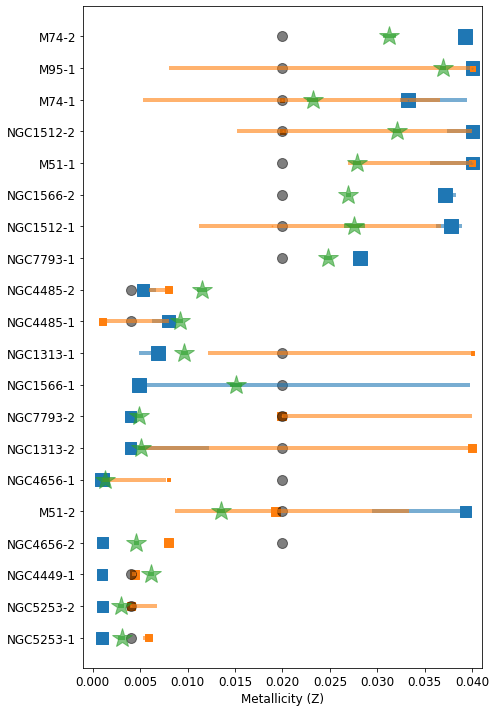}\\
	\includegraphics[width=0.40\textwidth]{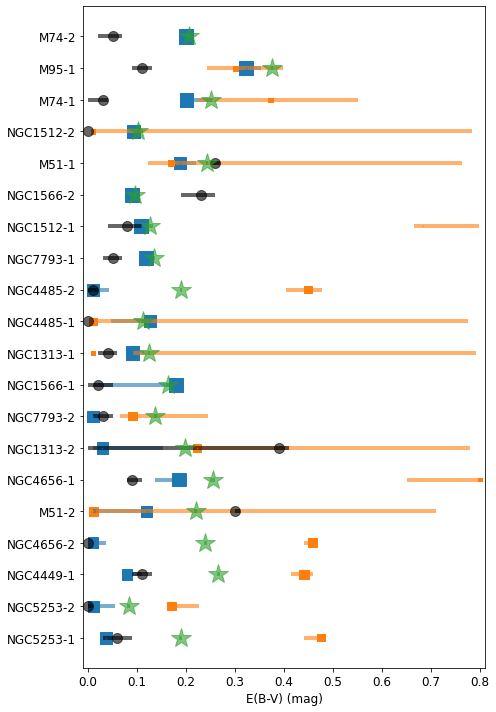}
	\includegraphics[width=0.40\textwidth]{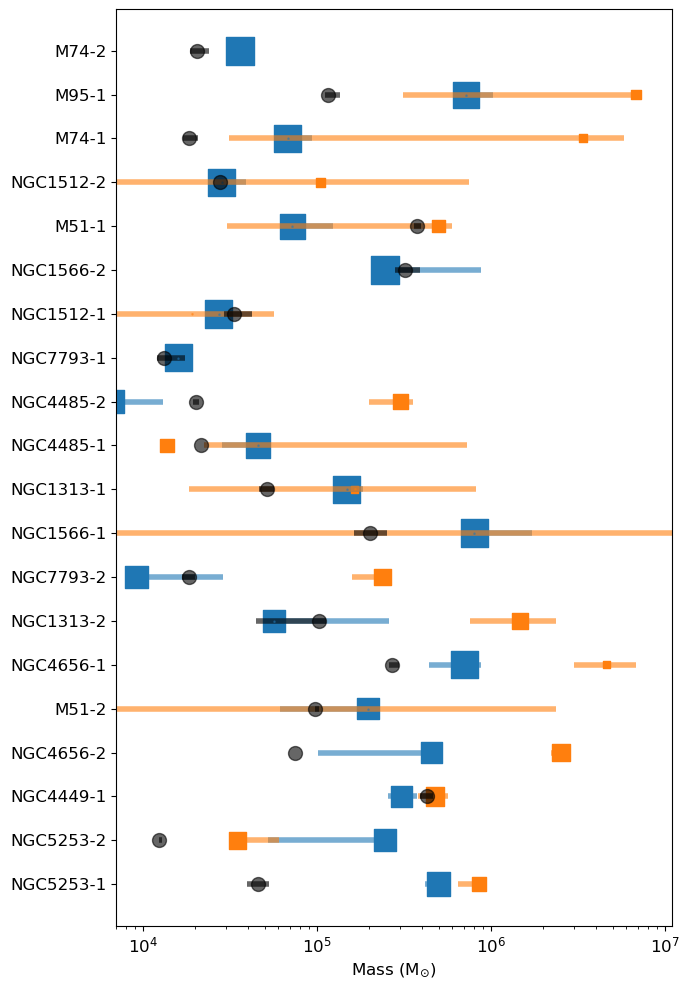}\\
    \caption{Values of the best-fit parameters for ages, metallicities, reddening attenuations and masses of our different models. The grey dots represent the values derived from photometry. The blue and orange squares are the values for \textit{Pop 1} and \textit{Pop 2} respectively, of our double population model of the FUV spectroscopy, with their size proportional to their light flux contribution at 1270 Å. The green stars are the average values of our multiple population model.}
    \label{fig:parameters}
\end{figure*}

Table \ref{tab:fit} lists the best-fit parameters for each stellar population of the CLUES sources compared with the values obtained with the other methods (multiple populations approach and photometric values). 
The uncertainties on the best-fit values are determined with a classic Monte Carlo approach and are shown in Figure \ref{fig:parameters} and Table \ref{tab:fit}.

\begin{figure*}
	\includegraphics[width=\textwidth]{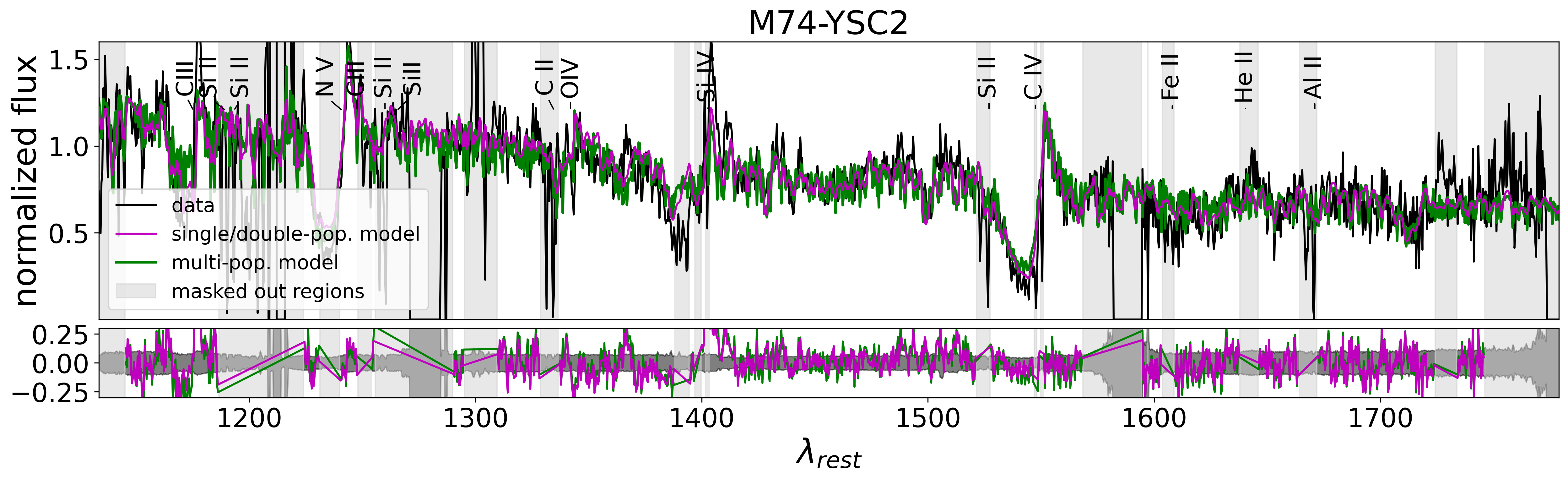}
	\includegraphics[width=0.34\textwidth]{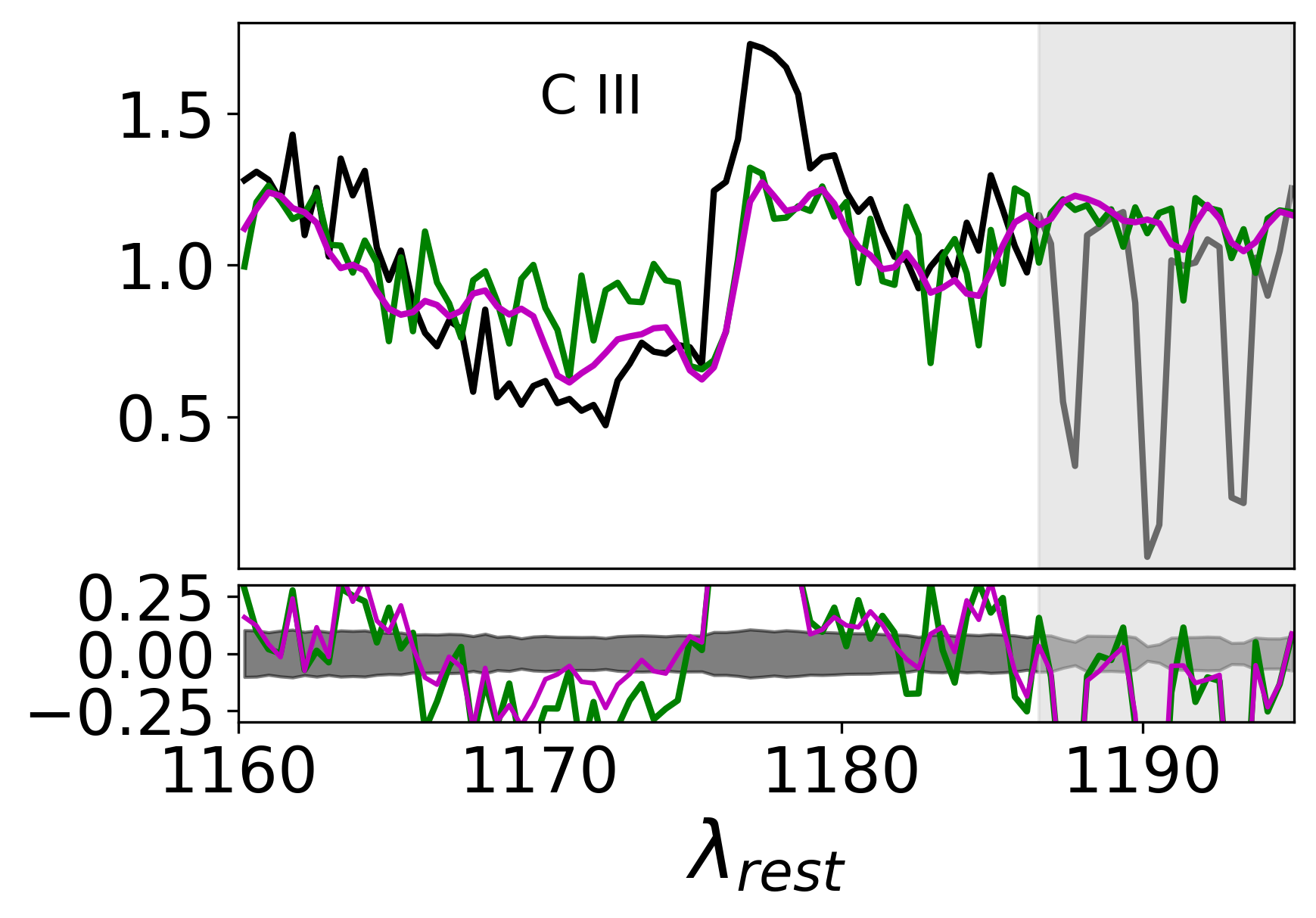}
	\includegraphics[width=0.31\textwidth]{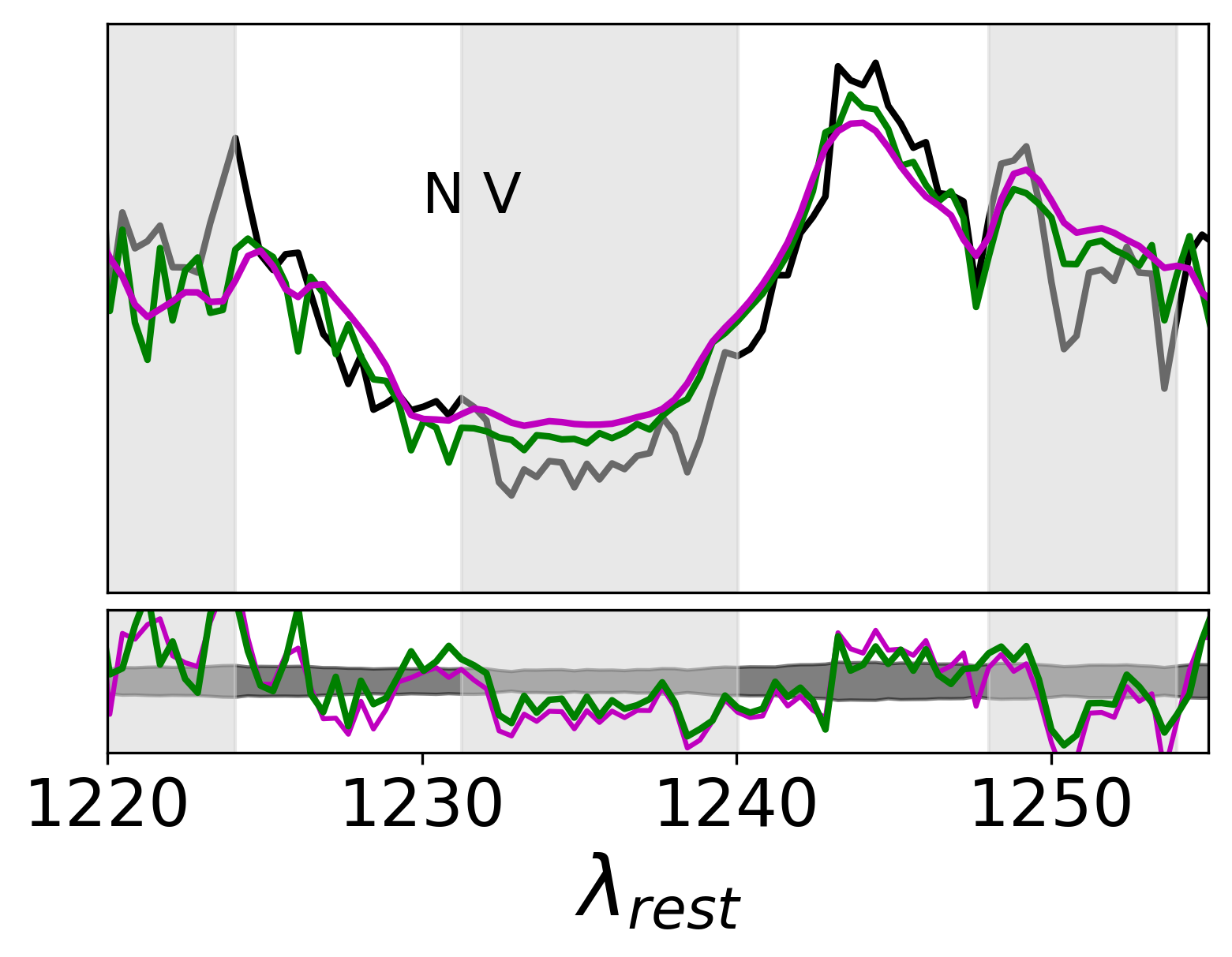}
	\includegraphics[width=0.31\textwidth]{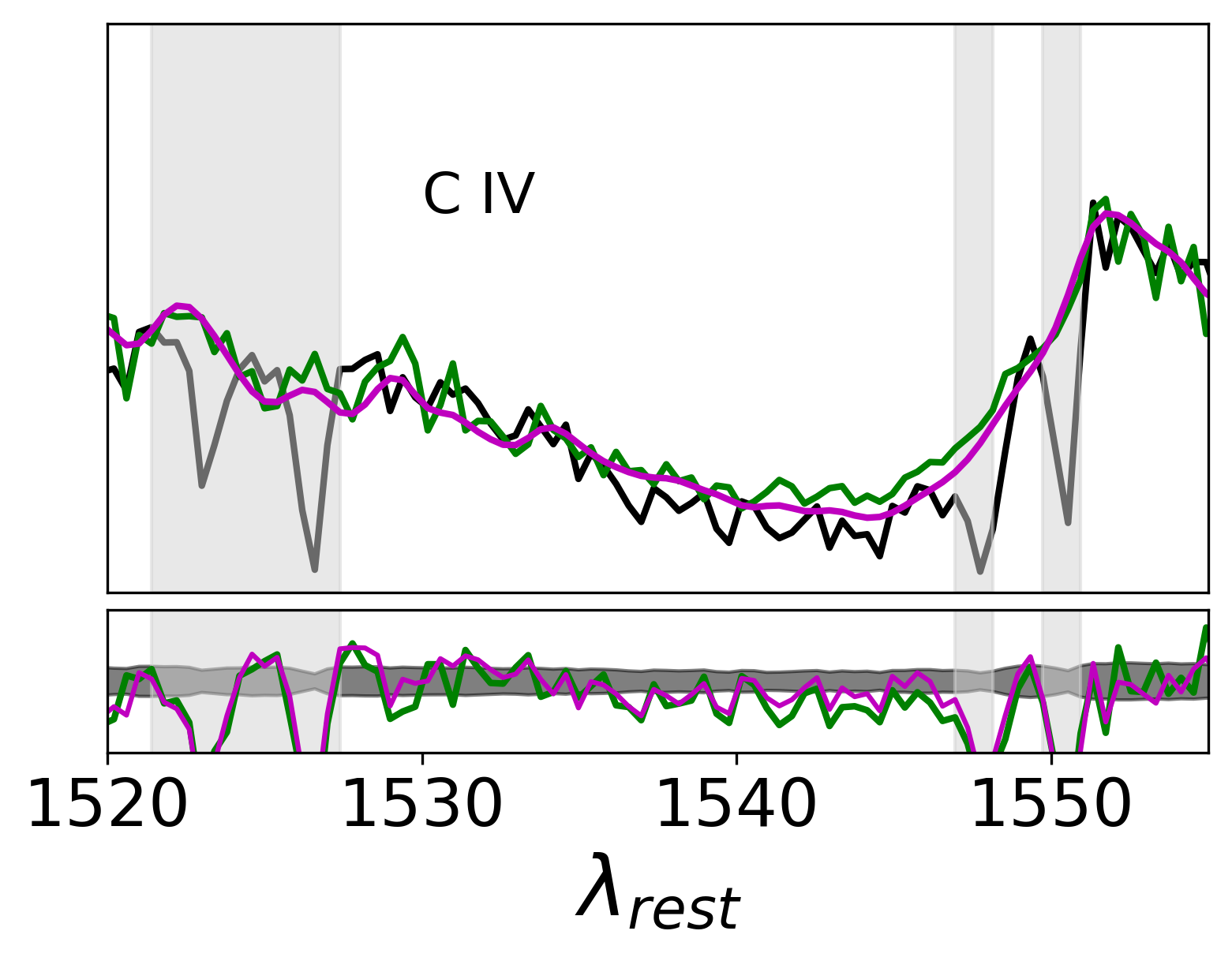}
	    
	\includegraphics[width=\textwidth]{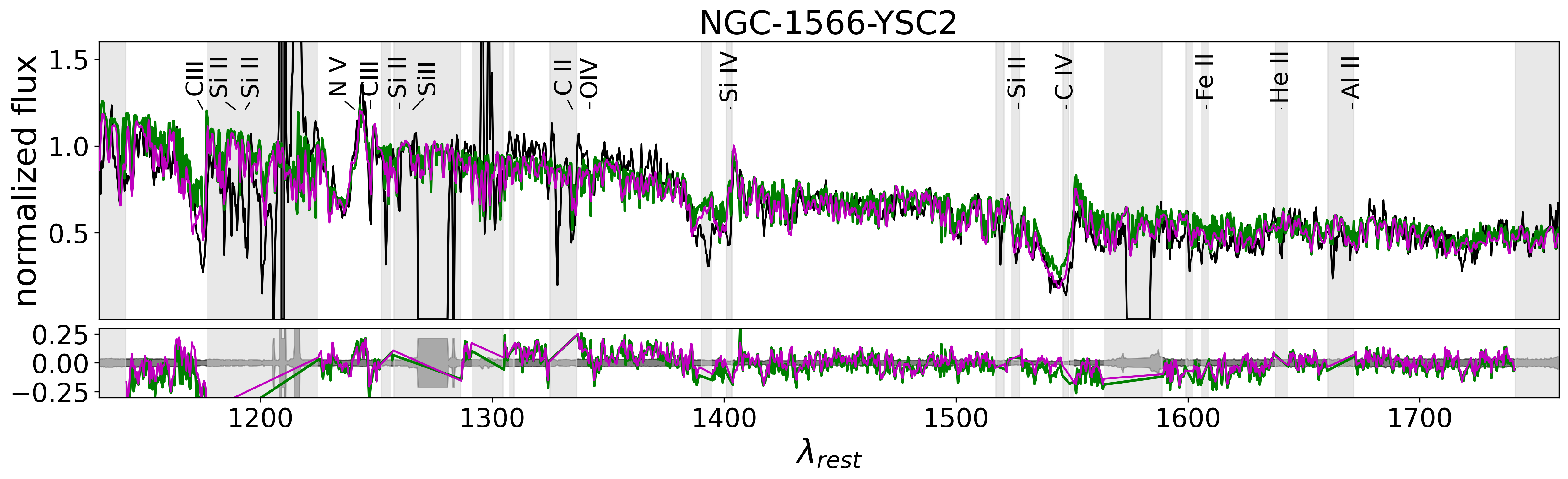}
	\includegraphics[width=0.34\textwidth]{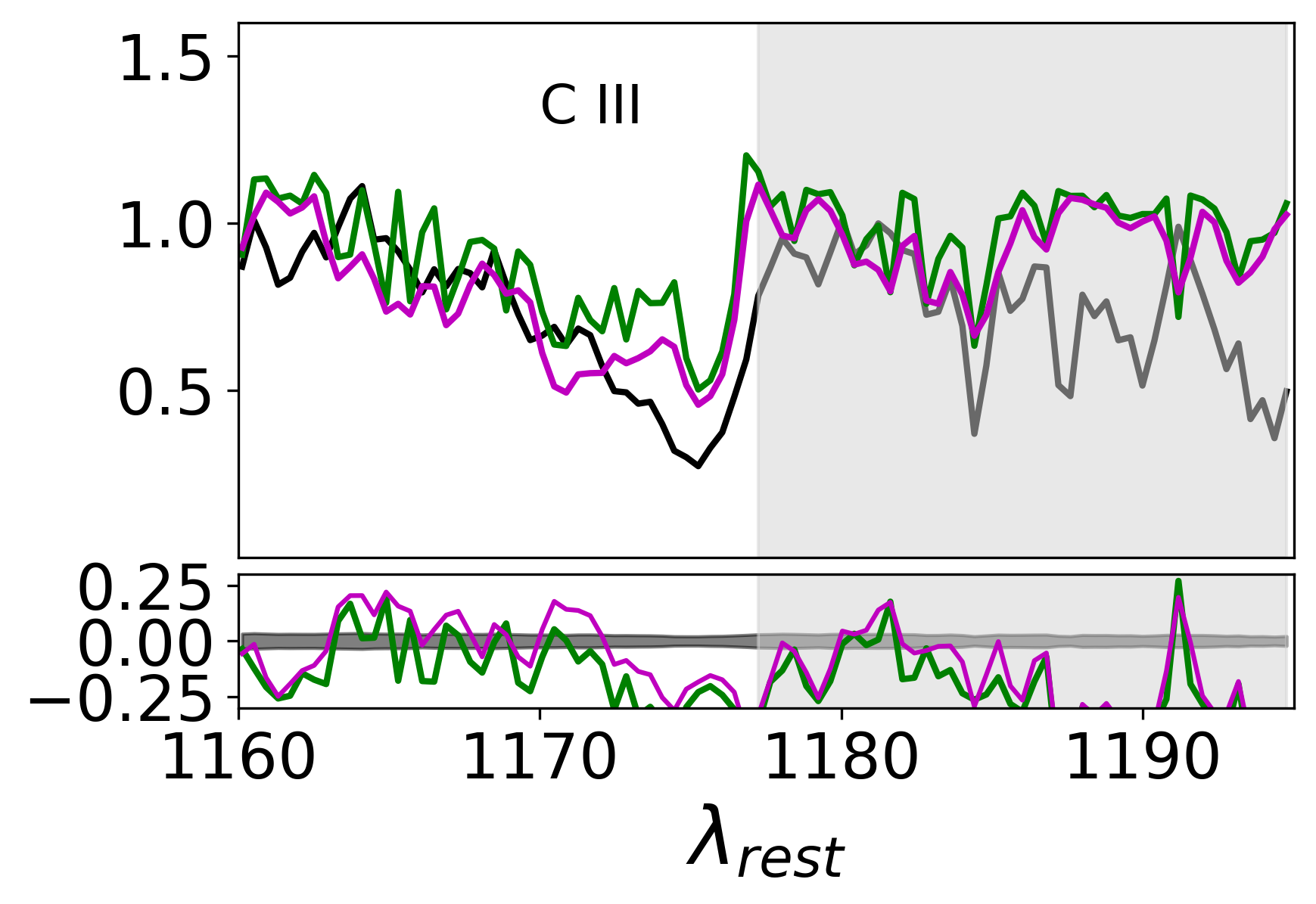}
	\includegraphics[width=0.31\textwidth]{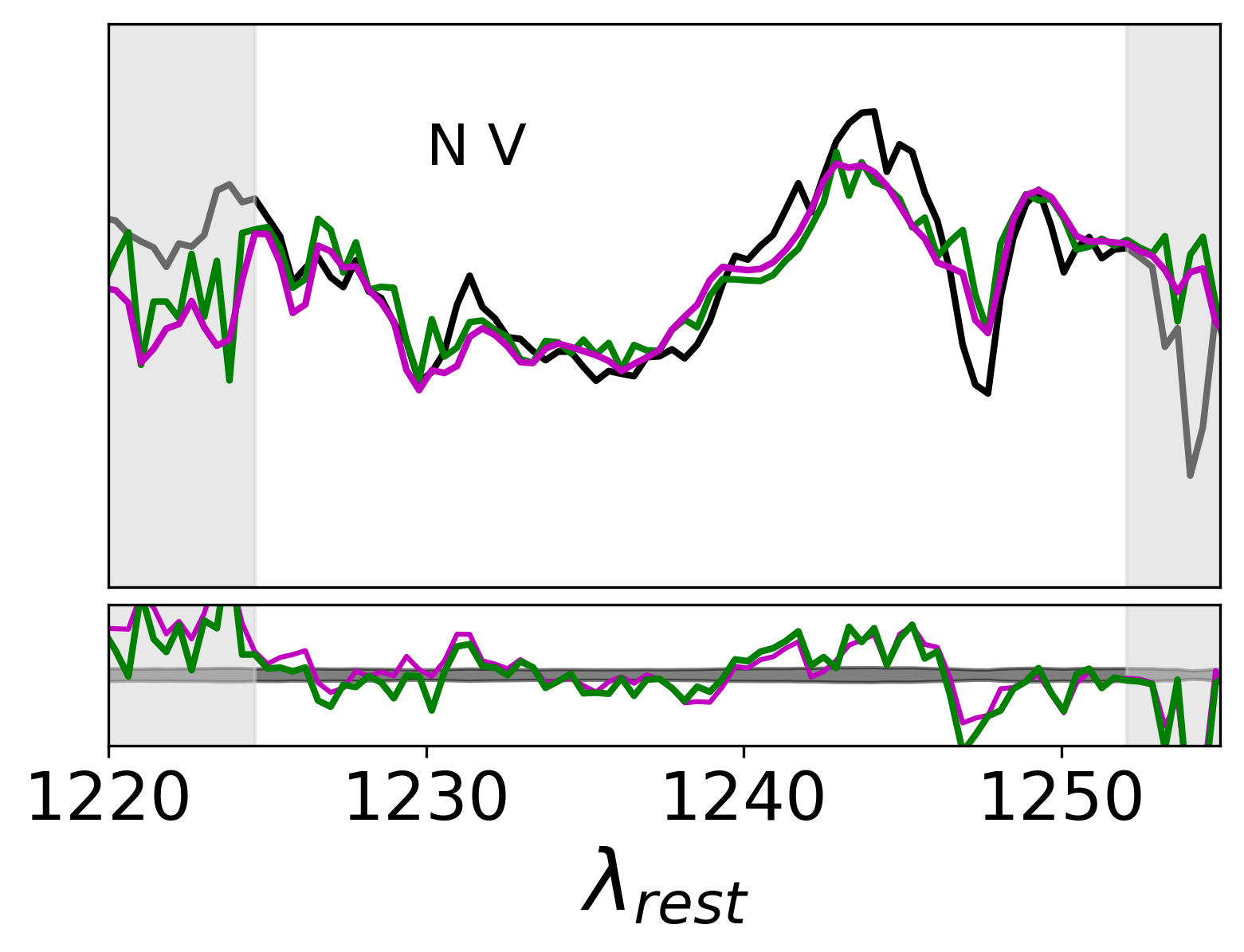}
	\includegraphics[width=0.31\textwidth]{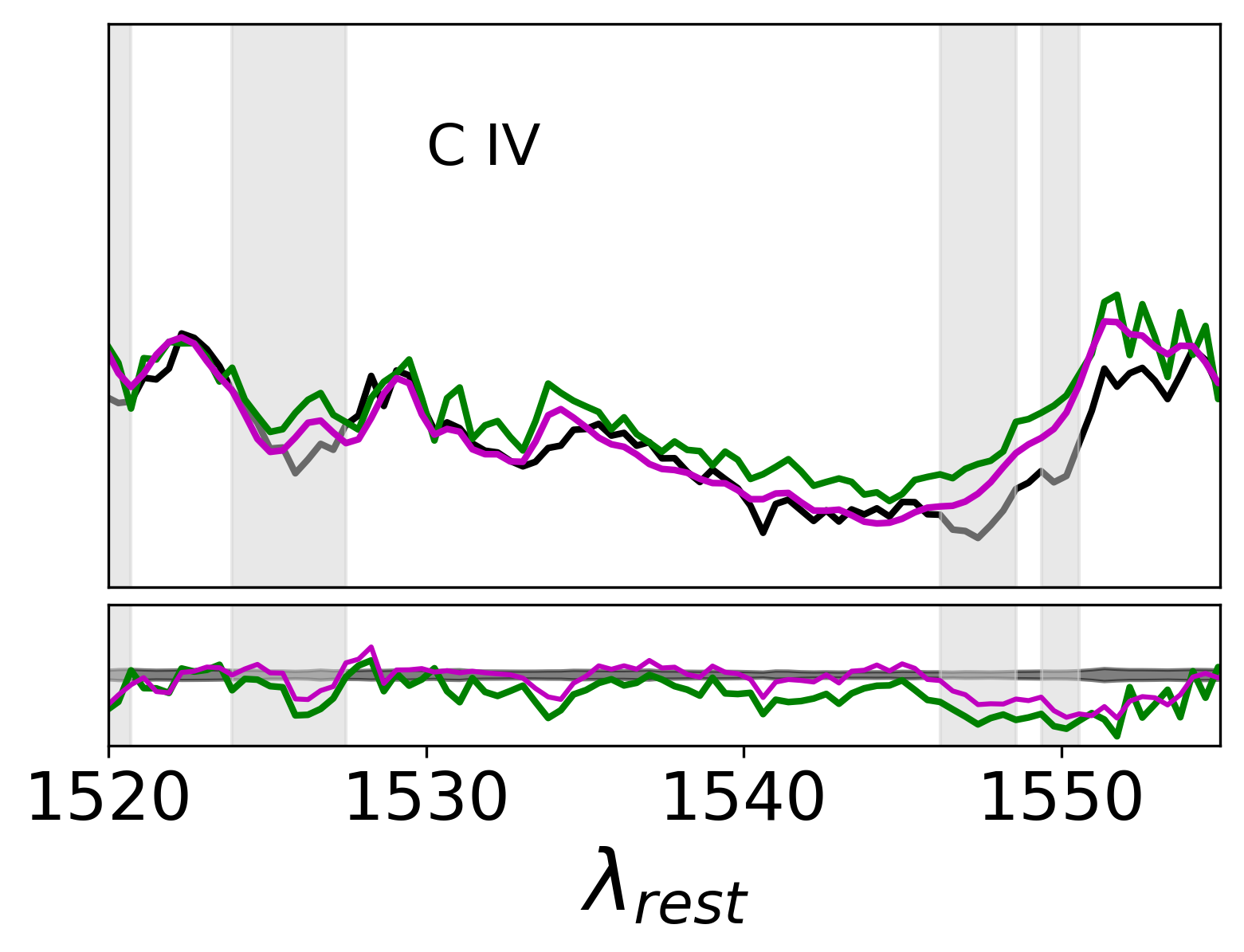}
	     
    \caption{Two targets (M-74-YSC2 and NGC-1566-YSC2) as examples of the fits to the FUV spectra (black line) with the single/double-population model and the multi-population model (purple and green lines respectively). The two panels below each target show a zoom into the P-Cygni lines C III, N V and C IV. The example targets chosen for this figure are a good and a poor agreement for the stellar age between single/double pop. model and multi-pop. model.}
    \label{fig:example}
\end{figure*}

\begin{table*}
\centering
\fontsize{8}{10}\selectfont
\begin{tabular}{lrrrrrrrcc}
\hline
 source    &   $age_{ph}$ (Myr) &    $age_1$ &    $age_2$ &   $<age>$ & $E(B-V)_{ph}$  &   $E(B-V)_1$ &   $E(B-V)_2$ &   $<E(B-V)>$\\
 (1) & (2) & (3) & (4) & (5) & (6) & (7) & (8) & (9)\\[0.1cm]
\hline
 M74-2     & 2.0$^{+0.5}_{-1.0}$  & 2.3$^{+0.0}_{-0.0}$    & -    & 2.3$^{+0.0}_{-0.0}$  & 0.05$^{+0.02}_{-0.03}$ & 0.20$^{+0.00}_{-0.00}$  & -     & 0.205$^{+0.005}_{-0.005}$ \\
 M95-1     & 2.0$^{+0.5}_{-1.0}$  & 2.5$^{+0.0}_{-0.0}$    & 42.0$^{+5.5}_{-40.1}$  & 2.8$^{+0.0}_{-0.0}$  & 0.11$^{+0.02}_{-0.02}$ & 0.32$^{+0.03}_{-0.00}$ & 0.30$^{+0.10}_{-0.06}$  & 0.376$^{+0.002}_{-0.002}$ \\
 M74-1     & 2.0$^{+0.5}_{-1.0}$  & 2.6$^{+0.1}_{-0.0}$    & 47.4$^{+1.9}_{-44.8}$  & 2.4$^{+0.0}_{-0.0}$  & 0.03$^{+0.01}_{-0.03}$ & 0.20$^{+0.05}_{-0.00}$  & 0.37$^{+0.18}_{-0.15}$  & 0.251$^{+0.002}_{-0.002}$ \\
 NGC1512-2 & 2.0$^{+0.5}_{-0.5}$  & 2.6$^{+0.1}_{-0.0}$    & 47.4$^{+2.1}_{-43.3}$  & 2.9$^{+0.0}_{-0.0}$  & 0.00$^{+0.00}_{-0.00}$ & 0.09$^{+0.02}_{-0.01}$  & 0.01$^{+0.77}_{-0.00}$  & 0.101$^{+0.002}_{-0.002}$ \\
 M51-1     & 2.0$^{+0.5}_{-0.5}$  & 3.4$^{+0.0}_{-0.0}$    & 35.7$^{+7.8}_{-16.5}$  & 3.6$^{+0.1}_{-0.1}$  & 0.26$^{+0.01}_{-0.01}$ & 0.19$^{+0.03}_{-0.01}$  & 0.17$^{+0.59}_{-0.05}$  & 0.242$^{+0.002}_{-0.002}$ \\
 NGC1566-2 & 2.0$^{+0.5}_{-1.0}$  & 3.4$^{+4.8}_{-0.0}$    & -    & 20.7$^{+1.1}_{-1.1}$ & 0.23$^{+0.03}_{-0.04}$ & 0.09$^{+0.00}_{-0.00}$  & -     & 0.163$^{+0.004}_{-0.004}$ \\
 NGC1512-1 & 2.0$^{+0.5}_{-1.0}$  & 3.5$^{+0.0}_{-0.0}$    & 47.4$^{+1.2}_{-41.4}$  & 3.1$^{+0.1}_{-0.1}$  & 0.08$^{+0.03}_{-0.04}$ & 0.11$^{+0.01}_{-0.01}$  & 0.68$^{+0.11}_{-0.02}$  & 0.126$^{+0.007}_{-0.007}$ \\
 NGC7793-1 & 3.0$^{+0.5}_{-1.0}$  & 3.5$^{+0.0}_{-0.0}$    & -    & 3.9$^{+0.0}_{-0.0}$  & 0.05$^{+0.02}_{-0.02}$ & 0.12$^{+0.00}_{-0.00}$  & -     & 0.135$^{+0.002}_{-0.002}$ \\
 NGC4485-2 & 4.0$^{+0.5}_{-0.5}$  & 3.7$^{+0.0}_{-0.1}$    & 2.1$^{+0.8}_{-0.0}$    & 3.3$^{+0.0}_{-0.0}$  & 0.01$^{+0.01}_{-0.01}$ & 0.01$^{+0.03}_{-0.00}$  & 0.45$^{+0.03}_{-0.04}$  & 0.190$^{+0.002}_{-0.002}$ \\
 NGC4485-1 & 4.0$^{+0.5}_{-0.5}$  & 4.4$^{+0.4}_{-0.0}$    & 14.1$^{+30.5}_{-12.1}$ & 5.8$^{+0.4}_{-0.4}$  & 0.00$^{+0.00}_{-0.00}$ & 0.13$^{+0.00}_{-0.08}$ & 0.01$^{+0.77}_{-0.00}$  & 0.111$^{+0.002}_{-0.002}$ \\
 NGC1313-1 & 1.0$^{+1.0}_{-0.5}$  & 4.9$^{+0.1}_{-0.1}$    & 36.1$^{+10.4}_{-30.9}$ & 6.7$^{+0.3}_{-0.3}$  & 0.04$^{+0.02}_{-0.02}$ & 0.09$^{+0.02}_{-0.00}$ & 0.01$^{+0.78}_{-0.00}$ & 0.125$^{+0.002}_{-0.002}$ \\
 NGC1566-1 & 11.0$^{+1.0}_{-1.0}$ & 8.0$^{+27.8}_{-0.0}$   & -    & 4.2$^{+0.1}_{-0.1}$  & 0.02$^{+0.03}_{-0.02}$ & 0.18$^{+0.00}_{-0.17}$ & -     & 0.096$^{+0.002}_{-0.002}$ \\
 NGC7793-2 & 14.0$^{+0.5}_{-0.5}$ & 9.9$^{+9.4}_{-0.3}$    & 47.3$^{+0.0}_{-11.0}$  & 26.3$^{+1.5}_{-1.5}$ & 0.03$^{+0.02}_{-0.02}$ & 0.01$^{+0.01}_{-0.00}$  & 0.09$^{+0.15}_{-0.02}$  & 0.137$^{+0.003}_{-0.003}$ \\
 NGC1313-2 & 2.0$^{+12.0}_{-0.5}$ & 29.8$^{+17.5}_{-6.9}$  & 36.1$^{+11.0}_{-34.2}$ & 28.9$^{+2.0}_{-2.0}$ & 0.39$^{+0.02}_{-0.39}$ & 0.03$^{+0.12}_{-0.02}$  & 0.22$^{+0.56}_{-0.00}$  & 0.197$^{+0.004}_{-0.004}$ \\
 NGC4656-1 & 30.0$^{+2.5}_{-2.5}$ & 30.1$^{+4.8}_{-0.0}$  & 4.7$^{+20.7}_{-1.3}$   & 23.2$^{+2.3}_{-2.3}$ & 0.09$^{+0.02}_{-0.01}$ & 0.19$^{+0.00}_{-0.05}$  & 0.80$^{+0.00}_{-0.15}$  & 0.254$^{+0.005}_{-0.005}$ \\
 M51-2     & 2.0$^{+0.5}_{-0.5}$  & 36.0$^{+0.4}_{-0.2}$   & 4.4$^{+32.0}_{-1.2}$   & 20.1$^{+1.7}_{-1.7}$ & 0.30$^{+0.01}_{-0.00}$ & 0.12$^{+0.00}_{-0.11}$ & 0.01$^{+0.70}_{-0.00}$ & 0.220$^{+0.004}_{-0.004}$ \\
 NGC4656-2 & 4.0$^{+0.5}_{-0.5}$  & 36.4$^{+0.0}_{-30.5}$ & 3.1$^{+0.1}_{-0.0}$    & 10.8$^{+1.3}_{-1.3}$ & 0.00$^{+0.00}_{-0.00}$ & 0.01$^{+0.03}_{-0.00}$  & 0.46$^{+0.01}_{-0.02}$  & 0.239$^{+0.003}_{-0.003}$ \\
 NGC4449-1 & 30.0$^{+2.5}_{-2.5}$ & 46.1$^{+4.0}_{-0.0}$   & 3.0$^{+0.0}_{-0.0}$    & 9.2$^{+1.1}_{-1.1}$  & 0.11$^{+0.02}_{-0.02}$ & 0.08$^{+0.03}_{-0.01}$  & 0.44$^{+0.02}_{-0.03}$  & 0.266$^{+0.003}_{-0.003}$ \\
 NGC5253-2 & 10.0$^{+0.5}_{-0.5}$ & 49.1$^{+0.0}_{-38.5}$  & 3.1$^{+0.3}_{-0.0}$    & 18.0$^{+0.8}_{-0.8}$ & 0.00$^{+0.01}_{-0.00}$ & 0.01$^{+0.04}_{-0.00}$  & 0.17$^{+0.06}_{-0.00}$  & 0.083$^{+0.003}_{-0.003}$ \\
 NGC5253-1 & 10.0$^{+0.5}_{-0.5}$ & 49.1$^{+0.1}_{-0.1}$   & 3.0$^{+0.0}_{-0.0}$    & 23.4$^{+1.1}_{-1.1}$ & 0.06$^{+0.03}_{-0.03}$ & 0.04$^{+0.01}_{-0.01}$  & 0.48$^{+0.01}_{-0.03}$  & 0.189$^{+0.003}_{-0.003}$ \\
\hline
\hline
  &   $M_{ph}$ ($10^6\,\msun$) &              $M_1$ &              $M_2$ &    $Z_{ph}$ &    $Z_1$ &        $Z_2$ &    $<Z>$  &   $Pop 1$ & $Pop 2$ \\
   & (10) & (11) & (12) & (13) & (14) & (15) & (16) & (17) & (18) \\[0.1cm]
\hline
 M74-2     & 0.02$^{+0.00}_{-0.00}$  & 0.04$^{+0.00}_{-0.00}$  & -   & 0.02  & 0.039$^{+0.001}_{-0.001}$  & -      & 0.031$^{+0.001}_{-0.001}$ & 1 & - \\
 M95-1     & 0.12$^{+0.02}_{-0.01}$  & 0.72$^{+0.31}_{-0.00}$  & 6.80$^{+0.00}_{-6.48}$ & 0.02  & 0.040$^{+0.000}_{-0.000}$ & 0.040$^{+0.000}_{-0.032}$  & 0.037$^{+0.000}_{-0.000}$ & 0.88 & 0.12 \\
 M74-1     & 0.02$^{+0.00}_{-0.00}$  & 0.07$^{+0.03}_{-0.00}$  & 3.37$^{+2.45}_{-3.34}$  & 0.02  & 0.033$^{+0.006}_{-0.001}$  & 0.020$^{+0.017}_{-0.015}$  & 0.023$^{+0.001}_{-0.001}$ & 0.92 & 0.08 \\
 NGC1512-2 & 0.03$^{+0.00}_{-0.03}$ & 0.03$^{+0.01}_{-0.00}$  & 0.10$^{+0.65}_{-0.10}$  & 0.02  & 0.040$^{+0.000}_{-0.003}$  & 0.020$^{+0.020}_{-0.005}$  & 0.032$^{+0.000}_{-0.000}$ & 0.91 & 0.09 \\
 M51-1     & 0.38$^{+0.02}_{-0.02}$  & 0.07$^{+0.05}_{-0.00}$  & 0.50$^{+0.10}_{-0.47}$  & 0.02  & 0.040$^{+0.000}_{-0.004}$  & 0.040$^{+0.000}_{-0.013}$  & 0.028$^{+0.000}_{-0.000}$ & 0.81 & 0.19 \\
 NGC1566-2 & 0.32$^{+0.07}_{-0.04}$  & 0.25$^{+0.63}_{-0.00}$  & -   & 0.02  & 0.037$^{+0.001}_{-0.001}$  & -      & 0.015$^{+0.001}_{-0.001}$ & 1    & -    \\
 NGC1512-1 & 0.03$^{+0.01}_{-0.00}$  & 0.03$^{+0.00}_{-0.00}$  & 0.02$^{+0.04}_{-0.02}$  & 0.02  & 0.038$^{+0.001}_{-0.002}$  & 0.019$^{+0.018}_{-0.008}$  & 0.028$^{+0.001}_{-0.001}$ & 1    & 0    \\
 NGC7793-1 & 0.01$^{+0.00}_{-0.00}$  & 0.02$^{+0.00}_{-0.00}$  & -   & 0.02  & 0.028$^{+0.001}_{-0.000}$  & -      & 0.025$^{+0.000}_{-0.000}$ & 1 & - \\
 NGC4485-2 & 0.02$^{+0.00}_{-0.00}$  & 0.01$^{+0.01}_{-0.00}$  & 0.30$^{+0.05}_{-0.10}$  & 0.004 & 0.005$^{+0.001}_{-0.000}$  & 0.008$^{+0.000}_{-0.002}$  & 0.012$^{+0.001}_{-0.001}$ & 0.72 & 0.28 \\
 NGC4485-1 & 0.02$^{+0.00}_{-0.02}$ & 0.05$^{+0.00}_{-0.02}$  & 0.01$^{+0.72}_{-0.00}$ & 0.004 & 0.008$^{+0.000}_{-0.002}$  & 0.001$^{+0.007}_{-0.000}$  & 0.009$^{+0.000}_{-0.000}$ & 0.75 & 0.25 \\
 NGC1313-1 & 0.05$^{+0.00}_{-0.01}$  & 0.15$^{+0.03}_{-0.02}$  & 0.17$^{+0.65}_{-0.15}$  & 0.02  & 0.007$^{+0.000}_{-0.002}$  & 0.040$^{+0.000}_{-0.028}$ & 0.010$^{+0.000}_{-0.000}$ & 0.94 & 0.06 \\
 NGC1566-1 & 0.20$^{+0.05}_{-0.04}$  & 0.80$^{+0.93}_{-0.00}$  & -   & 0.02  & 0.005$^{+0.035}_{-0.000}$ & -      & 0.027$^{+0.000}_{-0.000}$ & 1 & - \\
 NGC7793-2 & 0.02$^{+0.00}_{-0.00}$  & 0.01$^{+0.02}_{-0.00}$  & 0.24$^{+0.02}_{-0.08}$  & 0.02  & 0.004$^{+0.000}_{-0.000}$  & 0.020$^{+0.020}_{-0.000}$  & 0.005$^{+0.000}_{-0.000}$ & 0.64 & 0.36 \\
 NGC1313-2 & 0.10$^{+0.01}_{-0.06}$  & 0.06$^{+0.20}_{-0.00}$  & 1.47$^{+0.88}_{-0.71}$  & 0.02  & 0.004$^{+0.008}_{-0.000}$  & 0.040$^{+0.000}_{-0.035}$  & 0.005$^{+0.000}_{-0.000}$ & 0.65 & 0.35 \\
 NGC4656-1 & 0.27$^{+0.03}_{-0.01}$  & 0.70$^{+0.18}_{-0.26}$  & 4.59$^{+2.21}_{-1.58}$  & 0.02  & 0.001$^{+0.000}_{-0.000}$  & 0.008$^{+0.000}_{-0.007}$ & 0.001$^{+0.000}_{-0.000}$ & 0.94 & 0.06 \\
 M51-2     & 0.10$^{+0.00}_{-0.00}$  & 0.20$^{+0.00}_{-0.14}$  & 0.00$^{+2.35}_{-0.00}$ & 0.02  & 0.039$^{+0.001}_{-0.010}$  & 0.019$^{+0.014}_{-0.011}$  & 0.014$^{+0.000}_{-0.000}$ & 0.58 & 0.42 \\
 NGC4656-2 & 0.07$^{+0.00}_{-0.07}$ & 0.45$^{+0.00}_{-0.35}$ & 2.53$^{+0.31}_{-0.32}$  & 0.02  & 0.001$^{+0.000}_{-0.000}$ & 0.008$^{+0.000}_{-0.000}$  & 0.005$^{+0.000}_{-0.000}$ & 0.59 & 0.41 \\
 NGC4449-1 & 0.43$^{+0.04}_{-0.04}$  & 0.31$^{+0.07}_{-0.05}$  & 0.48$^{+0.09}_{-0.10}$  & 0.004 & 0.001$^{+0.000}_{-0.000}$  & 0.004$^{+0.000}_{-0.000}$  & 0.006$^{+0.000}_{-0.000}$ & 0.56 & 0.44 \\
 NGC5253-2 & 0.01$^{+0.00}_{-0.00}$  & 0.25$^{+0.00}_{-0.19}$  & 0.04$^{+0.03}_{-0.00}$  & 0.004 & 0.001$^{+0.000}_{-0.000}$  & 0.004$^{+0.003}_{-0.000}$  & 0.003$^{+0.000}_{-0.000}$ & 0.62 & 0.38 \\
 NGC5253-1 & 0.05$^{+0.01}_{-0.01}$  & 0.50$^{+0.05}_{-0.08}$  & 0.85$^{+0.10}_{-0.20}$  & 0.004 & 0.001$^{+0.000}_{-0.000}$ & 0.006$^{+0.000}_{-0.001}$  & 0.003$^{+0.000}_{-0.000}$ & 0.73 & 0.27 \\
\hline
\end{tabular}
\caption{Stellar population properties of the CLUES sample. (1) Star cluster name. (2) Age derived from photometry. (3) Age of \textit{Pop 1} of the double-population model for spectroscopy. (4) Age of \textit{Pop 2} of the double-population model for spectroscopy. (5) Light-weighted age of the multiple population model for spectroscopy. (6) $E(B-V)$ derived from photometry. (7) $E(B-V)$ of \textit{Pop 1} of the double-population model for spectroscopy. (8) $E(B-V)$ of \textit{Pop 2} of the double-population model for spectroscopy. (9) Light-weighted $E(B-V)$ of the multiple population model for spectroscopy. (10) Mass derived from photometry. (11) Mass of \textit{Pop 1} of the double-population model for spectroscopy. (12) Mass of the \textit{Pop 2} of the double-population model for spectroscopy. (13) Metallicity value fixed in photometry models. (14) Metallicity of \textit{Pop 1} of the double-population model for spectroscopy. (15) Metallicity of \textit{Pop 1} of the double-population model for spectroscopy. (16) Light-weighted metallicity of the multiple population model for spectroscopy. (17) Light flux fraction at 1270 Å of \textit{Pop 1} of the double-population model for spectroscopy. (18) Light flux fraction at 1270 Å of \textit{Pop 2} of the double-population model for spectroscopy.}
\label{tab:fit}
\end{table*}

\subsection{Multiple populations}
We also use a different approach to fit the FUV spectra. The method is developed and presented by \citet{chisholm19}. This method does not limit the fit to a number of populations established \emph{a-priori}, but it includes multiple single stellar populations in a linear combination of models.  We use SB99 models with ages of 1, 2, 3, 4, 5, 8, 10, 15, 20, 40~Myr and metallicities ($Z$) of 0.001, 0.004, 0.008, 0.02, 0.04. These ages and metallicities are chosen to probe the parameter space where the stellar continuum changes most dramatically in the SB99 models. This means that there are 50 total models to fit the observed data. We self-consistently add a nebular continuum to the SB99 models by putting each individual SB99 model into \textsc{cloudy} \citep{ferland17} using an ionization parameter logU of -2.5 and an electron density of 100~cm$^{-3}$. This should give a more accurate estimate of the nebular continuum than what is given by SB99, although, at FUV wavelengths, the nebular component of the models employed is much smaller than the total flux (a few percents). The \citet{reddy16} attenuation law is used to redden the model spectra and determine a single best-fit $E(B-V)$ for all stellar components. This law is similar in shape to the \cite{calzetti2000} law but with a different normalization. This multiple population approach assumes that the FUV light is approximated as a linear combination of single-age and single-metallicity bursts of star formation. We fit for the fraction of the total light that each model contributes using \textsc{mpfit} \citep{mpfit}. We then calculated light-weighted ages, metallicites and reddening attenuations (listed in Table \ref{tab:fit}) using the fitted light fractions. The errors for the multiple population fits are determined using a Monte Carlo approach. First we modulate each flux element by a Gaussian with a width equal to the error of the flux measurement and centered on zero. We then refit the modulated flux spectra and tabulate the age, metallicity, and $E(B-V)$. We repeat this process 100 times and take the standard deviation of the distribution as the error on the measurements (Table \ref{tab:fit}). The median age and metallicity relative error is 0.042 and 0.038, respectively, for the full sample. 

\section{Results}
\label{sec:results}

\subsection{FUV spectral fitting: comparison between the two methods}

The variations in the P-Cygni lines are clearly visible in Figure~\ref{fig:atlas} where the clusters are sorted by increasing age of \textit{Pop 1}. The strength of the P-Cygni lines decreases with increasing age, acting as the strongest age-sensitive features in the spectra at this range of ages.

In Figure~\ref{fig:example} we show two examples of FUV cluster spectra fitted by the two methods (the remaining cluster spectral fitting analysis is included in the appendix). There are no significant differences between the two model spectra (double-population in magenta and multi-population in green), suggesting that both models perform similarly. In general, we note that P-Cygni lines such as C IV and N V are well fitted by the models, while C III remains poorly constrained (e.g., bottom panels of Figure~\ref{fig:example}). The continuum slope of the observed spectra is also well reproduced by the models. 

We notice that we do not fit the broad He II wind feature, which arises in the presence of strong and fast stellar winds typical of very massive stars \citep[VMS, e.g.][]{Smith2016} and Wolf-Rayet phases because, even though they are included in the models, they are not often populated in the evolutionary tracks. 

In our sample, two clusters show strong He II emission. In the case of NGC4485-YSC2 the He II emission is very broad (see Figure~\ref{fig:app2}). For this cluster, all methods produce an age of 3-4 Myr, i.e. compatible with Wolf-Rayet stellar evolution phases. The other cluster showing strong He II emission is M74-YSC2 (see Figure~\ref{fig:example}), the youngest cluster of the sample, with a best recovered age of 2.3 Myr. 
This age implies that the cluster is too young to have entered the WR phase. One solution is that the He II emission arises in very massive stars, although we consider this unlikely for the following reasons. Strong Si IV P-Cygni emission is present in the M74-YSC2 spectrum but this feature is absent in VMS spectra \citep{Crowther2016, Smith2016} because it is associated with cooler and older OB supergiants. A key spectral diagnostic of VMS is the presence of O V, indicating that the hottest and youngest O stars are present, and O V is not detected with any certainty in M74-YSC2. VMS have also not yet been found at the supersolar metallicity of M74-YSC2. We thus prefer to explain the presence of He II emission at 2.3 Myr by arguing that the massive stars in the cluster have already entered the classic WR phase because of the high O star mass-loss rates at supersolar metallicity. 
We note that the scaling of O star mass loss rates with $Z$ above Z$_\odot$, as used in the models, is not well known and has not been observationally tested. Rotational mixing may also be important and this, in combination with the enhanced O star mass loss rates, could potentially lead to an earlier WR phase. However, this possibility is slightly in tension with the latest Geneva grid of models at supersolar metallicities \citep[][]{Yusof2022}, which find the evolution to the WR phase starting at 3 Myr for rotating models. Observations of very young (2-3 Myr) massive star clusters at high metallicities would definitively help to further investigate this important stellar evolutionary phase. 

 Finally, several clusters are in the age range 3-4 Myr but do not show HeII in emission suggesting they might be younger (they have all strong P-Cygni stellar features supporting their young age.)

\subsection{Best-fit parameters of the stellar populations}
In Figure~\ref{fig:parameters}, we display the recovered physical parameters of the CLUES clusters by using the diverse approaches listed above. The clusters are sorted by increasing age of the \textit{Pop 1}, as determined by the two-population methodology.

Overall the photometric SED fitting results (grey dots in Figure~\ref{fig:parameters}) show typical trends already recognised and reported in the literature especially for young star clusters (ages below $\sim10$Myr) \citep[e.g.][]{wofford2016, hannon2022}. We notice that clusters with ages below 3 Myr have a quite large uncertainty (up to a few Myr), implying that optical photometric SEDs are quite insensitive below this age. At ages between 7 and 20 Myr, large uncertainties are produced due to the degeneracy in the color space produced affecting red supergiants (e.g. Figure~\ref{fig:selection}), between ages and extinction that are in some cases reflected in the recovered uncertainties of the cluster physical parameters (e.g. NGC1313-YSC2) and not in others (e.g. NGC4646-YSC2). Another important factor to note is that metallicity is not fitted but assumed in photometric SED analyses, a limitation well known in the literature \citep[e.g.][]{anders2004}.



 For the two-population fit (blue and orange squares in Figure~\ref{fig:parameters}) we notice that \textit{Pop 1} has well-constrained parameters, whereas \textit{Pop 2} is the model with the largest errors and its parameters are often unconstrained. In Table~\ref{tab:fit}, we list what fraction of each population dominates the total flux at 1270 \AA. \textit{Pop 1} by definition contributes more than 50\% of the observed flux of the spectrum. The inclusion of \textit{Pop 2} improves the quality of the fit in 16 out of the 20 clusters based on the AIC criterion. However, the associated uncertainties to the physical parameters in 9 of these 16 clusters suggest that the physical parameters of this second stellar population remain unconstrained in some of the clusters. We will discuss in Sec \ref{sec:phot} that this is likely because this second population represents a mix of stellar populations, not necessarily formed in a single burst, included in the COS aperture. In the aperture of the older clusters (e.g. NGC4656-YSC2, NGC4449-YSC1, NGC5253-YSC2, NGC5253-YSC1) we detect a second population of young stars, 3 Myr old, necessary to account for the weak P-Cygni lines (see Figure \ref{fig:app3}). We note that these recently formed stars have a higher metallicity than the older clusters in the same regions.  

The average age values of the multiple-population analysis (the green stars in Figure \ref{fig:parameters}) show overall very good agreement with the two-population method. In 11 of the 20 clusters, the average flux-weighted ages coincide within 1$\sigma$ with the ones of \textit{Pop 1} (blue squares). This agreement is largely driven by the fact that this population dominates the flux at the reference wavelength. In the remaining clusters, the multiple-population solutions are bracketed between the best values recovered for \textit{Pop 1} and \textit{Pop 2}. In these cases, the contribution of \textit{Pop 2} (orange squares) is between 20 and 40\% of the flux, hence significantly weighting on the resulting multiple-population method. 

We find that the best-fit ages derived from photometry are consistent within the uncertainties with the values derived from spectroscopy only in about half of the targets. Ten of the 20 clusters have photometric age between 1 and 2 Myr and uncertainties from a few to several Myr. Spectroscopic ages range between 2 and 4 Myr for 7 of these clusters, partially overlapping within the uncertainties with the photometric ones, but with significantly narrower uncertainties. For the other 3 clusters with very young photometric ages, the FUV spectroscopy produces ages between 5 and 20 Myr. Indeed the spectra of these clusters do not show typical P-Cygni features of massive stars, which suggests that they are significantly older than the estimated photometric ages. 

Four of the clusters with photometric ages between 3 and 4 Myr, are within the uncertainties of the recovered ages of \textit{Pop 1} and \textit{Pop 2}. In general we see that as we move to older clusters, the discrepancies between ages from photometry and ages from spectroscopy increase, suggesting that the photometric derived values are degenerate. We see the changes in the FUV spectral features that reinforce these results, such as weaker P-Cygni lines.

Metallicity is not fitted in the photometric analysis, while it is a free parameter in the FUV spectral fitting. We warn the reader that the metallicity values of \textit{Pop 2} are largely unconstrained as can be seen by the wide error bars in Figure \ref{fig:parameters}. This is due to the degeneracy of the model and the fact that the model variation can be very small for a certain variation of the metallicity, relative to other parameters. The average metallicity of the multiple stellar population model with discrete values of ages and metallicites (green stars in Figure \ref{fig:parameters}) is typically in between the values for \textit{Pop 1} and \textit{Pop 2} (blue and orange squares). We believe this average value is the closest to the true metallicity, especially when there is a large difference between \textit{Pop 1} and \textit{Pop 2} that cannot be physical given the vicinity of the two stellar populations (e.g. NGC1313-1, NGC5253-1, NGC4485-1 and NGC4656-2). Overall we note that FUV spectral fitting produces metallicity values which are closer to solar or super-solar values for large spirals (M74, M51, NGC1566, M95). Deviations are observed for the two spirals NGC1313 and NGC7793, where FUV spectral analyses suggest significantly lower metallicities. These results are in agreement with nebular measurement abundances reported in the literature for both galaxies \citep[e.g.][]{walsh1997, dellabruna2021} which support sub-solar metallicities. In the case of M-51, we see in Figure \ref{fig:parameters} that the metallicity of YSC1, which is closer to the galaxy centre, is higher than that of YSC2, satisfying the expectation of an inside-out star formation in the galaxy. On the other hand, for NGC-1512 the metallicities of YSC1 and YSC2 are similar but the low S/N ratio of the spectroscopy of this target may make the results uncertain.

Extinction is also a key parameter for photometric SED fitting analysis. In general we see that uncertainties are relatively small in the extinctions derived from photometry, except when the uncertainties in the ages are significant (e.g., see NGC1313-YSC2). We observe that the extinction derived from optical analysis is significantly lower than that obtained from FUV spectroscopy of the youngest clusters, suggesting that FUV light might be more affected by reddening which these young star clusters suffer. In general, extinction has large uncertainties in the \textit{Pop 2} recovered values, suggesting indeed that this second population is not well constrained. For slightly older clusters, the extinction derived for \textit{Pop 1} agrees quite well with photometric values and multiple-population fit. We highlight that when the \textit{Pop 2} has a noticeable contribution to the cluster light (e.g., NGC4656-YSC2, NGC4449-YSC1, NGC5253-YSC1 and YSC2) the extinction recovered by the multiple-population fit is bracketed between the values of the two populations.\footnote{The multiple-population fit uses the \cite{reddy16} extinction attenuation which has the same functional scape of the \cite{calzetti2000} but different normalisation. In Figure~\ref{fig:parameters} we plot E(B-V), which is a differential value, thus we do not expect to see differences.} 

In the majority of the cases, spectroscopically derived masses are significantly larger (from a factor of 2 to several tens) than those obtained from photometry. This discrepancy is expected because the COS aperture and the photometric region from which photometry of the cluster is extracted do not match, with the former being larger. We also note that, since by construction \textit{Pop 1} is the major light contributor, its resulting mass is lower than \textit{Pop 2}, which appears to be on average older and, therefore, heavier (higher mass-to-light ratio). 


\subsection{Spatially resolving the stellar and cluster light within the COS aperture}
\label{sec:phot}
In this section, we take advantage of the spatial information that the LEGUS imaging provides of each CLUES region to investigate what fraction of the light that enters the COS aperture is dominated by the cluster light and whether the colors would support the presence of distinct stellar populations.

 We perform aperture photometry of the CLUES targets using two different circular apertures: one with 1.25" radius equal to the COS aperture that includes the total flux as in the FUV spectra, the other with 0.2" radius that includes only or mostly the flux of the star cluster. In both cases we apply a local sky-background subtraction using an annulus of 1.5" radius and width of 0.08".  Fluxes are then converted into Vega-mag, using the tabulated LEGUS zeropoints. In this way we are able to differentiate the flux contribution from the star cluster and the one from the diffuse stars, for each HST filter. We take into account the COS vignetting by weighting the flux of each pixel within the COS aperture by the COS sensitivity function. The central part of the COS aperture is more sensitive than the outer parts and therefore the light of the central cluster contributes to a larger extent in the COS spectra.
 
\begin{figure*}
    \centering
    \includegraphics[width=0.9\textwidth]{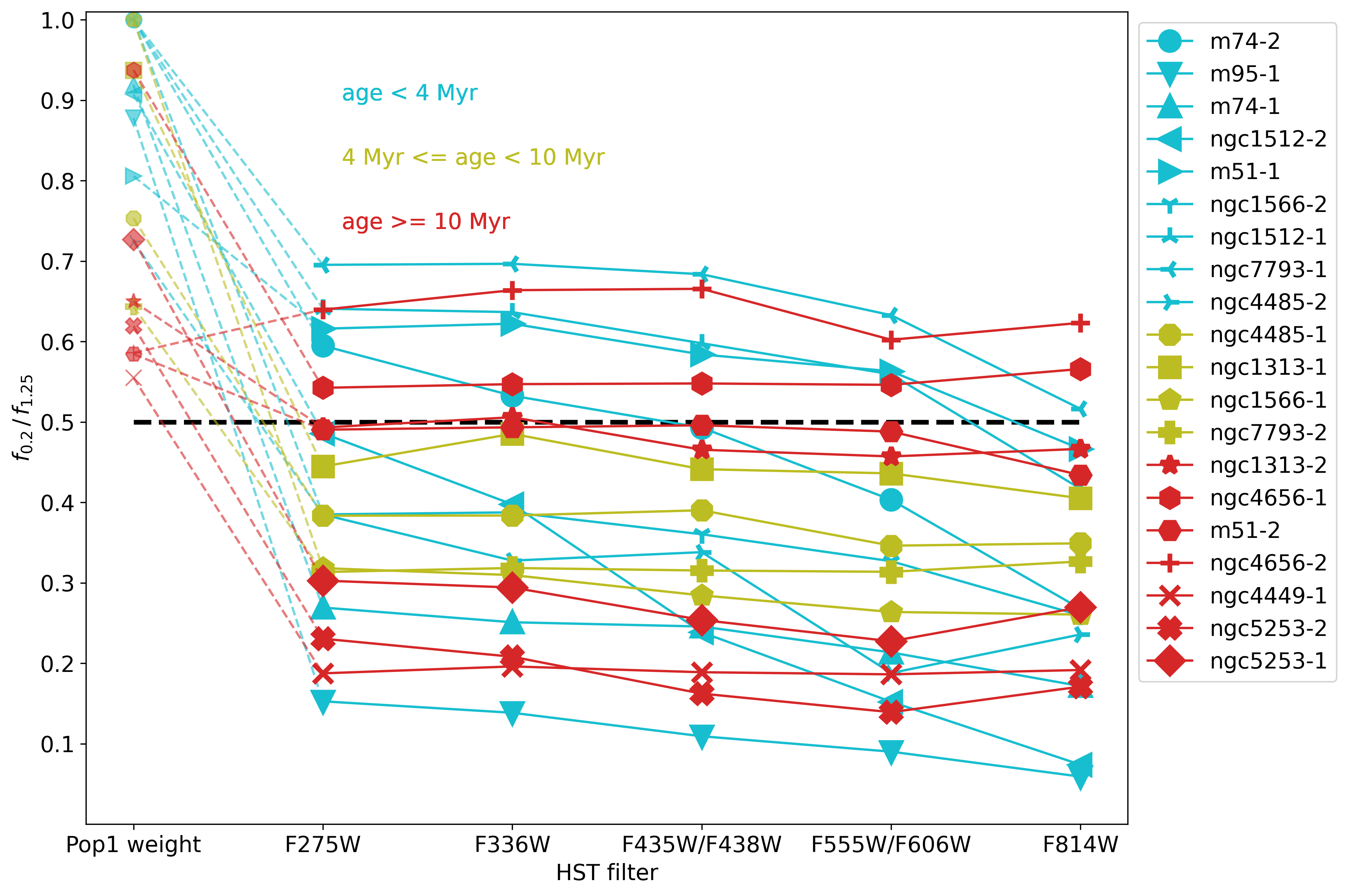}
    \includegraphics[width=0.7\textwidth]{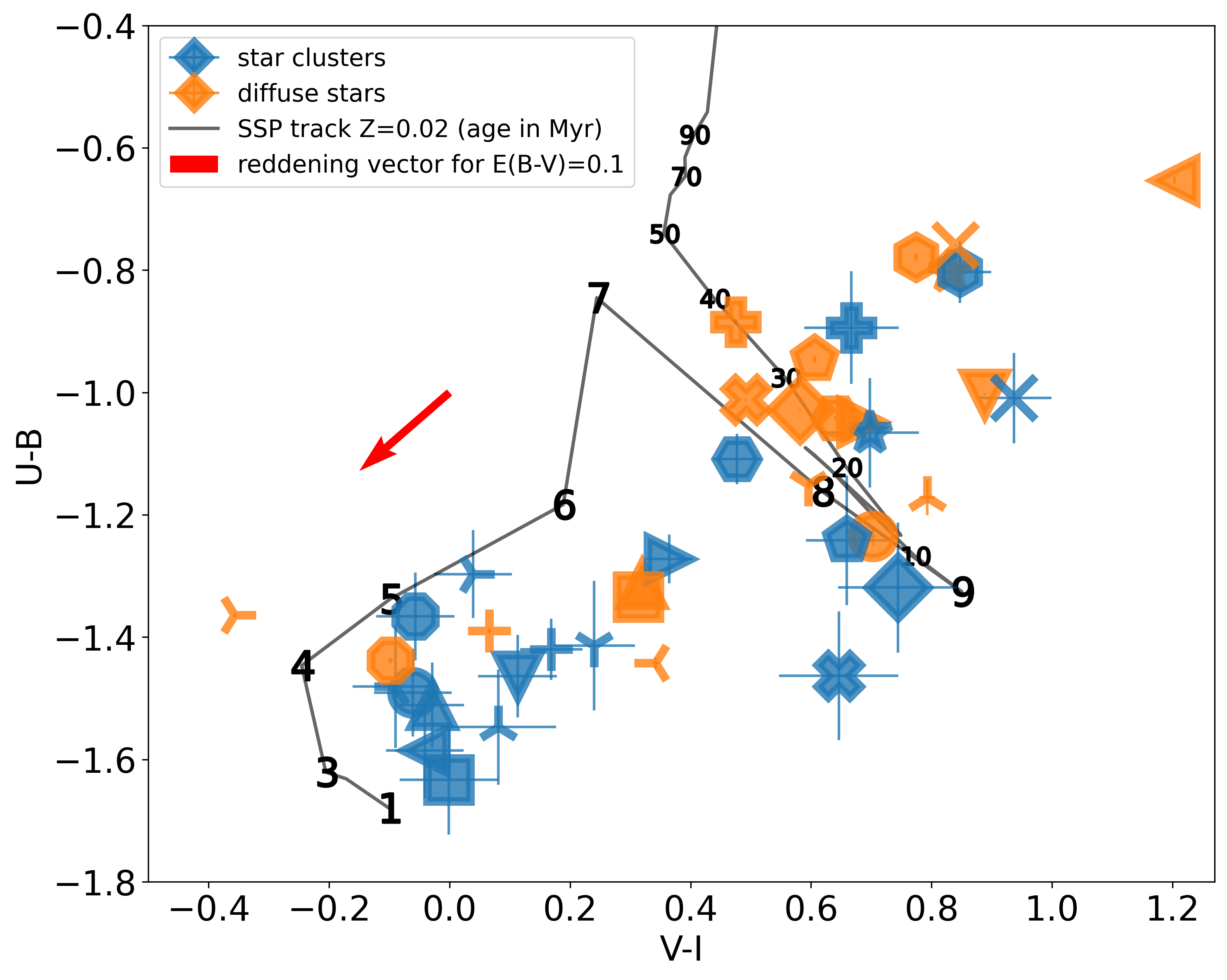}
     \caption{(Top panel) Ratio of light flux from the star cluster vs. the total flux within the COS aperture, for each of the HST filter indicated on the x-axis as well as for the fictitious FUV filter (Pop 1 weight, see text in Section \ref{sec:phot}). Each CLUES source is represented with a different marker. The colors of the markers and the lines divide the sources in three groups according to the age of \textit{Pop 1}: $<$4 Myr (cyan), 4 - 10 Myr (olive green), $\geq$ 10 Myr (red). (Bottom panel) Color-color plot $V-I$ vs. $U-B$ for both the light flux from the central star cluster (0.2" radius aperture) marked in blue and the diffuse stars (outside the 0.2" radius aperture and within the COS aperture) marked in orange. The black line marks the stellar track of a single stellar population with Z=0.02, with the main age steps printed in black along the line. The reddening vector for $E(B-V)$=0.1 is shown in red.}
     \label{fig:phot_ratios}
\end{figure*}

 In Figure \ref{fig:phot_ratios}, top panel, we show the ratio between the flux from the cluster (at 0.2" radius, $f_{0.2}$) and the total flux within the COS aperture (1.25" radius, $f_{1.25}$), as a function of the HST filter. The first column of this panel, \textit{Pop1 weight}, is the fraction of light flux at 1270 Å contributed by the first population relative to the total light of the double-population model as recovered from the double-population fit. This can be seen as a fictitious FUV filter. We note that the COS spectroscopy is taken in the FUV portion of the spectrum, whereas we extend here tour analysis to optical wavelengths in order to investigate the presence of multiple stellar populations by looking at the trends of the flux ratios over a large range of wavelengths.
 
 Overall, in the top panel of Figure \ref{fig:phot_ratios}, we observe that only about 30\% of the CLUES targets have 50\% or more of their NUV light (F275W) coincident with the star cluster. In 4 of these targets (M74-YSC2, NGC4656-YSC1, NGC-1512-YSC1 and NGC7793-YSC1) we recover from the FUV fit listed in Table~\ref{tab:fit}, that the contribution of \textit{Pop 2} to the total light at 1270 Å is zero or up to only few per cent, i.e. the cluster dominates the light within the aperture. In the other 2 clusters (M51-YSC1 and NGC4656-YSC2) the best fit produces a \textit{Pop 2} contributing between 19\% and 41\% of the FUV light. In the four targets where a single-population model was sufficient to describe the observed spectroscopy (M74-YSC2, NGC7793-YSC1, NGC1566-YSC1, NGC1566-YSC2), there is no need to model an extra component of stellar light in addition to the central cluster. This can happen for three reasons: (i) there is actually no significant stellar component around the central cluster and within the COS aperture (e.g., see image of M74-YSC2 in Figure \ref{fig:images}); (ii) there is an additional stellar component but it has similar age, metallicity and reddening attenuation to the principal one so that the two are indistinguishable (e.g. NGC7793-YSC1): (iii) the second population is significantly older, therefore redder, with an insignificant contribution in the FUV.
 
If we limit our analysis to the LEGUS NUV-to-NIR bands, we notice that the fraction of light in the clusters for which the FUV spectral fitting produces ages younger than 4 Myr, decrease from the FUV to the NIR band of at least 10 \% (and up to a maximum of 40\%). This behaviour is not observed in the intermediate and older age clusters, where the changes are limited to only a few per cent. This trend could be explained if star formation in the youngest star cluster regions has not propagated in the surroundings but it is concentrated in proximity of the cluster. The background disk stellar population of the region where the cluster is forming is contributing to the diffuse light within the COS aperture and significantly contributing at longer wavelengths. As time goes on, star formation propagates in the entire region entering the COS aperture. The ratio of the cluster light with respect to the diffuse light remains constant as it is dominated by the recently formed stars throughout the region.   
 In general we see that \textit{Pop 1} dominates the light in the FUV. This is especially true for very young star clusters (cyan symbols) where at least 70\% of the light appears to be produced by very young stars ($<$4 Myr). The contribution of the intermediate and older age clusters spans a significantly larger range. Moreover the difference between the \textit{Pop1 weight} and the F275W ratio (on average lower than 50\%) confirms that the latter is still heavily affected by stellar populations in the range 5-30 Myr (as shown by our spectroscopic analysis).

We further investigate the color of the clusters and diffuse surrounding regions in Figure \ref{fig:phot_ratios}, lower panel, with the goal of confirming the presence of a diffuse stellar population contributing to the COS FUV spectra. 
We observe that all the clusters that have a spectroscopy-based \textit{Pop1} age younger than 4 Myr, have optical colors that are close to the model color corresponding to ages between 1 and 6 Myr (after accounting for extinctions). The other clusters with resulting \textit{Pop1} age between 5 and 50 Myr are all located in proximity of the red-supergiant phase in the evolutionary tracks, again suggesting that the FUV spectral fit produces reliable ages. In the case of the color of the diffuse population, we see that the regions around clusters NGC4485-YSC1 and YSC2, NGC4656-YSC2, NGC1313-YSC1, M74-YSC1, NGC7793-YSC1 have colors that overlap with the youngest star clusters, suggesting also very young ages. Indeed, when cross-matching these regions with the output ages of \textit{Pop2}, we see that the FUV spectroscopy produces in all the cases (except for NGC7793-YSC1 which is best fitted by a single population) either young ages or old ages but with large uncertainties encompassing a range from 3 to 50 Myr. We also notice that the majority of the clusters that correspond to \textit{Pop1} with ages below 4 Myr have diffuse light with redder $U-B$ and $V-I$ colors pointing toward an older stellar population surrounding the cluster.   

In summary we find that a significant fraction of the flux included in the COS aperture is associated with diffuse stellar population around the main massive cluster in the centre. By studying the colors of the clustered and diffuse stellar emission we confirm the cluster ages derived from spectroscopy and we highlight the degeneracy problem encountered with the photometry model.


\section{Discussion}
\label{sec:discuss}


We derived physical properties of star clusters with different methods based on both UV-to-optical high-spatial-resolution imaging and FUV spectroscopy from 1130 Å to 1770 Å (rest-frame) and compared the results. We found that multiple stellar populations are often needed to have the full picture of the star formation history of a certain region in the galaxy. Combining spectroscopy with photometry allows us to better interpret the nature of the multiple stellar populations within a star-forming region. This type of analysis that combines optical imaging and FUV spectroscopy will be feasible at high redshifts using JWST capabilities. A big improvement in spectroscopic depth of FUV studies at high-redshift will come when the Extremely Large Telescope (ELT) will start operating. In fact, for galaxies at redshift 2 to 4 the FUV restframe is redshifted to the optical.
    
\begin{figure}
        \centering
        \includegraphics[width=\columnwidth]{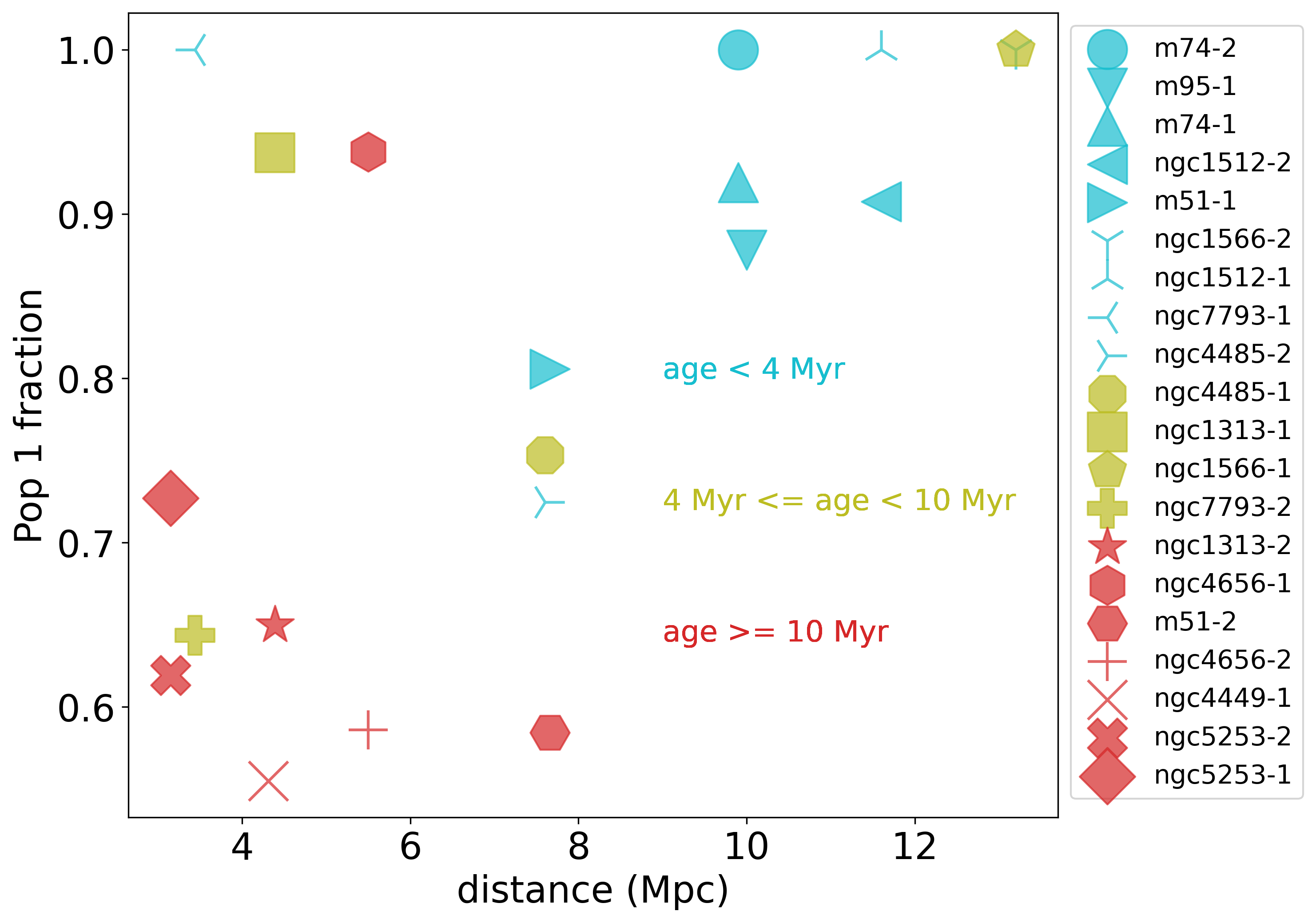}
        \caption{Weight of the FUV-dominant stellar population as a function of the distance of the target galaxy, color coded with the age of the stellar population. A different marker is used for each source, as listed in the legend.}
        \label{fig:distance}
    \end{figure}

Since the aperture from which we extract the FUV spectroscopy is fixed (COS aperture is 2.5" in diameter), the physical scale of each source changes as a function of the distance of the source, listed in Table \ref{tab:sample}. In Figure \ref{fig:distance} we plot the contribution of the FUV-dominant stellar population as a function of the distance to each source. While at distances closer than 8 Mpc, we find stellar populations that contribute to various degrees and at various ages, for the 7 sources at a distance larger than 8 Mpc the light contribution of \textit{Pop1} is larger than 85\%, a trend that may be due to selection effects of the CLUES sources. We also notice that the majority of clusters at distance larger than 8 Mpc are very young, which is consistent with them dominating the FUV flux, e.g., at larger distances only the brightest UV clusters are detectable. We repeated the same analysis replacing the age with the mass of \textit{Pop1} and we found a large spread of masses for these young clusters that dominate the FUV flux. This indicates that the young clusters dominate the FUV flux truly because of their young age rather than a large mass.
    
While high-redshift FUV studies infer the properties of stellar populations and the star formation history of star-forming galaxies \citep[e.g.][]{chisholm19}, the advantage of our analysis consists of the access to high spatial resolution imaging, which enables us to disentangle the clustered star formation from the diffuse stellar component. By comparing different methodologies in our analysis we highlight both the strengths and weaknesses of FUV spectroscopy as a tool to study young stellar populations. FUV spectroscopy remains one of the only tools we can use to break the age-extinction-metallicity degeneracy that affects optical photometry, in particular for clusters older than 5 Myr. However, we find it challenging to constrain the metallicity of the targets in our sample, possibly due to the low S/N ratio. Our best value is given by the average of multiple stellar populations, which are the same models used in \cite{chisholm19}. Here the authors show that the derived stellar metallicities are in agreement with the gas phase nebular metallicities for a sample of low-redshifts galaxies.
Considering the points above, we can conclude that overall FUV spectroscopy is a powerful tool to derive the physical properties of young star-forming regions. However, the FUV light is more affected than visible light by reddening, which is often the case for very young star clusters. 


The limitations in studying the FUV largely come from limitations of the models that not only have typically lower spectral resolution but also lack the implementation of certain phenomena that are common in massive stars and might play a role in shaping the FUV spectra, e.g. binary and multiple systems of stars, rotating stars \citep[][]{Grasha2021}, formation of very massive stars exceeding the upper limit of the IMF, presence of Wolf-Rayet stars. More sophisticated models that include the binary nature of massive stars will be studied in a future paper.

We noted in Section \ref{sec:results} that in our double-population models \textit{Pop 2} has typically larger uncertainties and hence unconstrained. This indicates a limitation of the model but it also suggests that the two different components of the model have a different nature. In this picture, \textit{Pop 1} is the stellar population of the central cluster formed of stars with similar ages, metallicities, intrinsic reddening; \textit{Pop 2} represents the diffuse stars included in the COS aperture or the smaller clusters around the central cluster, and therefore shows a larger range of ages, metallicities, reddenings. In this scenario, because the errors are estimated using a Monte Carlo simulation, the errors of a certain parameter of \textit{Pop 2} can be interpreted as a local spread of the galaxy for that physical parameter.

Finally we note that our analysis lacks any stochasticity effect in the stellar population synthesis employed. \cite{krumholz2015} showed that drawing stochastically the stars from the IMF enables more accurate deductions of the properties of young star clusters. However, since our selection of clusters consist of massive clusters ($10^4$ to $10^6 \rm\,\msun$), we are in a regime where these stochastic effects are minimal and we do not expect to see a significant difference should we use stochastic methods.

\section{Conclusions}
\label{sec:conclusions}

We present new COS FUV spectroscopy combined with existing  HST UV/optical photometry for a sample of 20 young, massive nearby star clusters selected from the LEGUS survey.

We present the physical properties of the stars populating young star clusters and their immediate surroundings, in a circular region between 38 and 160 pc across. The target clusters are hosted by nearby star-forming galaxies and are bright FUV sources. In our work we study both the FUV light, measured with the COS spectrograph, and the UV and optical light measured with the HST camera using different filters. Our analysis is based on different methodologies (i.e. photometric modelling and spectroscopic modelling) in order to measure the properties of the stellar population. We found that the majority of the star-forming regions of our sample are populated by more than one stellar population. Except in four cases (M-74-YSC2, NGC-7793-YSC1, NGC-1566-YSC1, NGC-1566-YSC2), at least two stellar population models are necessary to reproduce the FUV spectroscopy observed. We interpret one of the stellar populations (the one dominating the light flux in the FUV) to be the stellar component of the central star cluster in the region, and all the rest as diffuse stars in the surrounding as well as smaller star clusters that fall within the COS aperture. 

We measure ages for the youngest population in each region between 2 and 5 Myr with high accuracy ($< 1$ Myr), however the older stellar population(s) have larger uncertainties and are often unconstrained. The younger star clusters have higher metallicities ($Z\sim0.040$), whereas the older star clusters have lower metallicities ($Z\sim0.005$). This is mostly a consequence of the galaxies in which the clusters were selected. The intrinsic reddenings of the clusters range between 0.01 and 0.4, the secondary stellar component (diffuse stars) has typically larger reddening (up to 0.8) and larger uncertainties. The masses of the stellar populations that we studied range between $10^4$ and $10^7\,\msun$.

We conclude that FUV spectroscopy and stellar population synthesis are powerful tools to study the properties of young star clusters and star-forming galaxies. This kind of analysis gains even more insights when combined with photometry and spatially resolved data. This and similar works will benefit in the future by more complete and sophisticated models that include all the key physical phenomena and provide an output model spectrum with a higher resolution, matching the observations.

\acknowledgments

A. Adamo and M.S. acknowledge support from Swedish National Space Agency (SNSA) under the grant Dnr158/19.
A. Adamo acknowledges the support of the Swedish Research Council, Vetenskapsr\aa{}det. M.M. acknowledges the support of the Swedish Research Council, Vetenskapsrådet (internationell postdok grant 2019-00502).
K.G. is supported by the Australian Research Council through the Discovery Early Career Researcher Award (DECRA) Fellowship DE220100766 funded by the Australian Government.
K.G. is supported by the Australian Research Council Centre of Excellence for All Sky Astrophysics in 3 Dimensions (ASTRO~3D), through project number CE170100013. 
A.Aloisi and S.O. are grateful for the support for this program, HST-GO-15627, that was provided by NASA through a grant from the Space Telescope Science Institute, which is operated by the Associations of Universities for Research in Astronomy, Incorporated, under NASA contract NAS5- 26555. 
This project has received funding from the European Research Council (ERC) under the European Union’s Horizon 2020 research and innovation program (grant agreement number 757535)
C.U. acknowledges the support of the Swedish Research Council, Vetenskapsrådet.

%

\vspace{5mm}
\facilities{HST(COS)}


\software{astropy \citep{2013A&A...558A..33A},  
          Cloudy \citep{2013RMxAA..49..137F}, 
          SExtractor \citep{1996A&AS..117..393B}
          }

\appendix


\begin{figure*}

	\includegraphics[width=\textwidth]{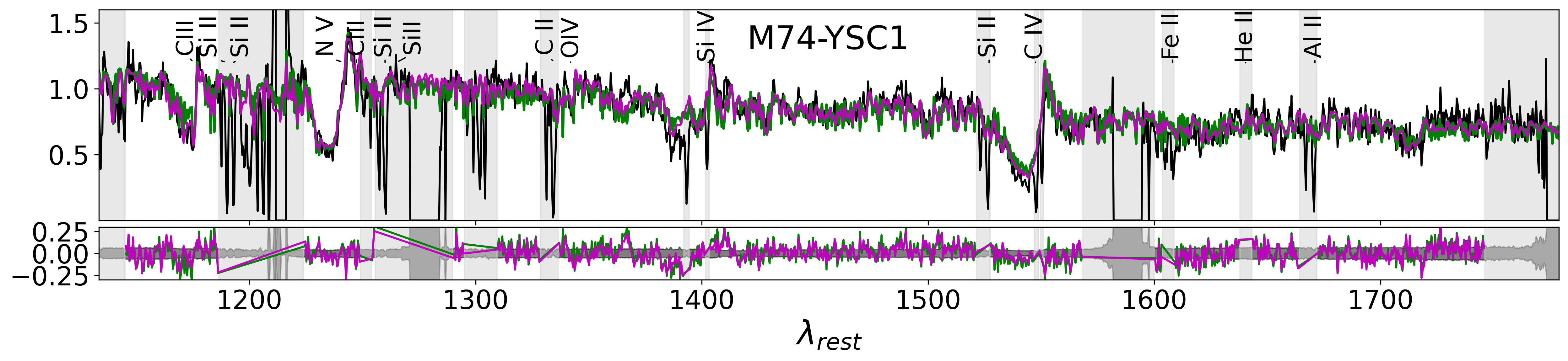}\\
	    
	\includegraphics[width=\textwidth]{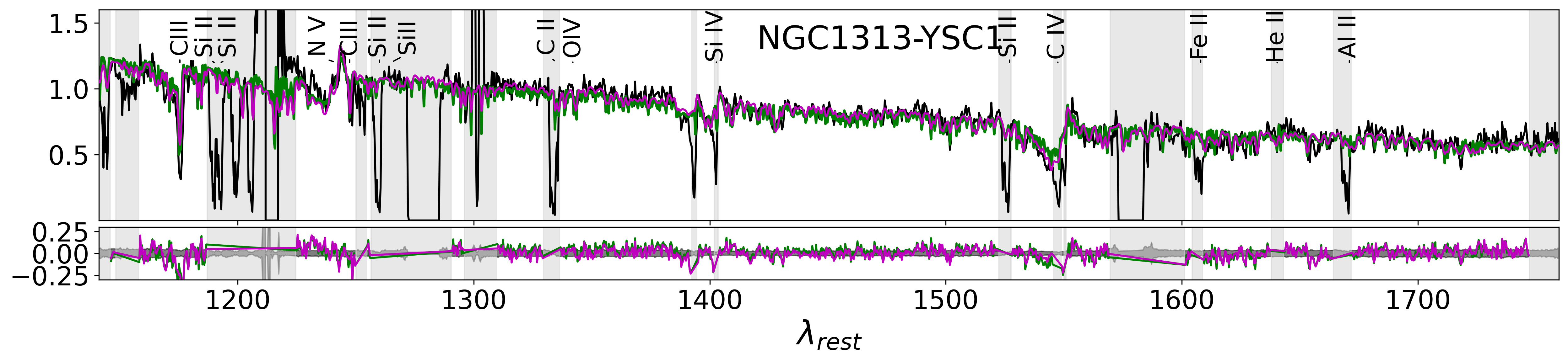}\\
	    
	\includegraphics[width=\textwidth]{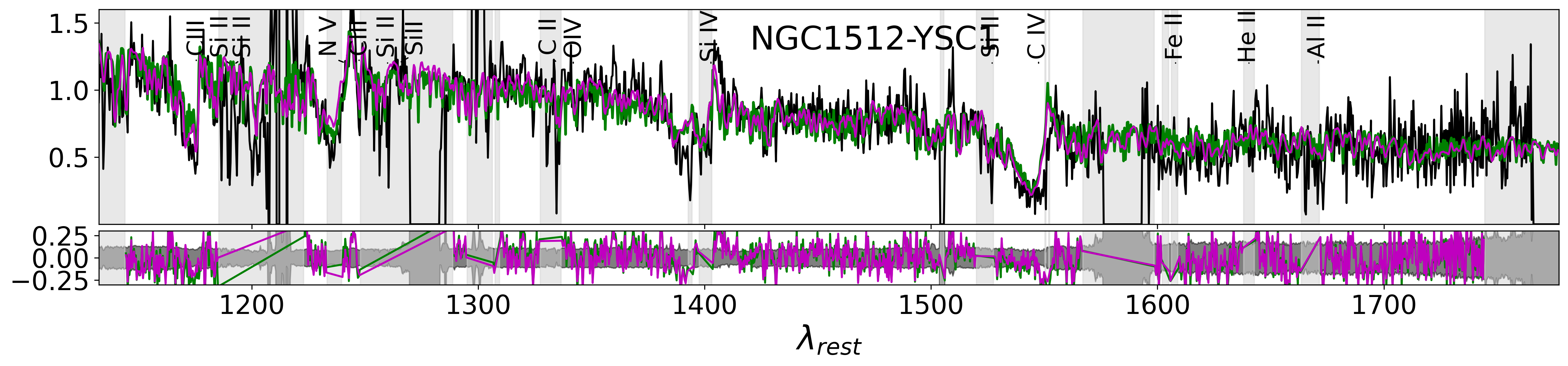}\\
	      
	\includegraphics[width=\textwidth]{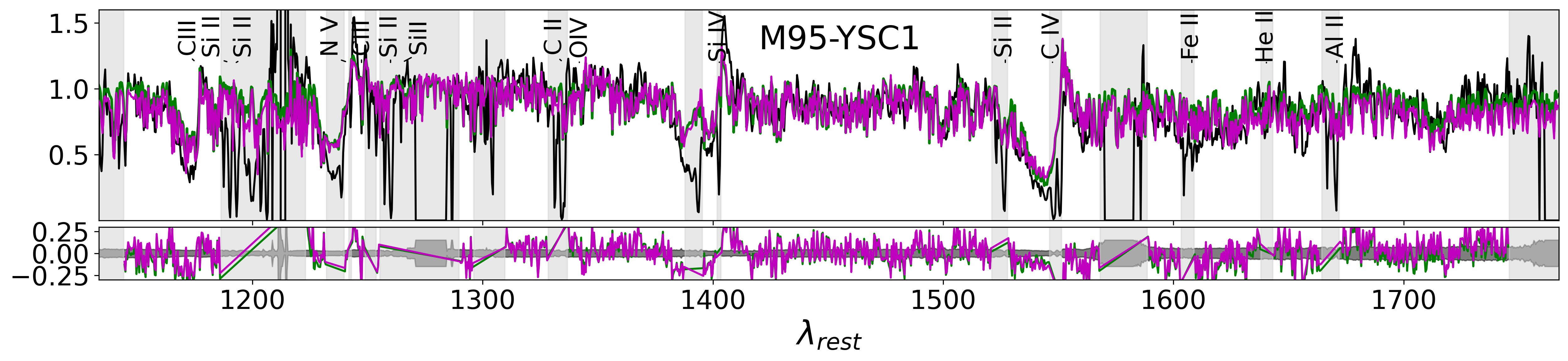}\\
	    
	\includegraphics[width=\textwidth]{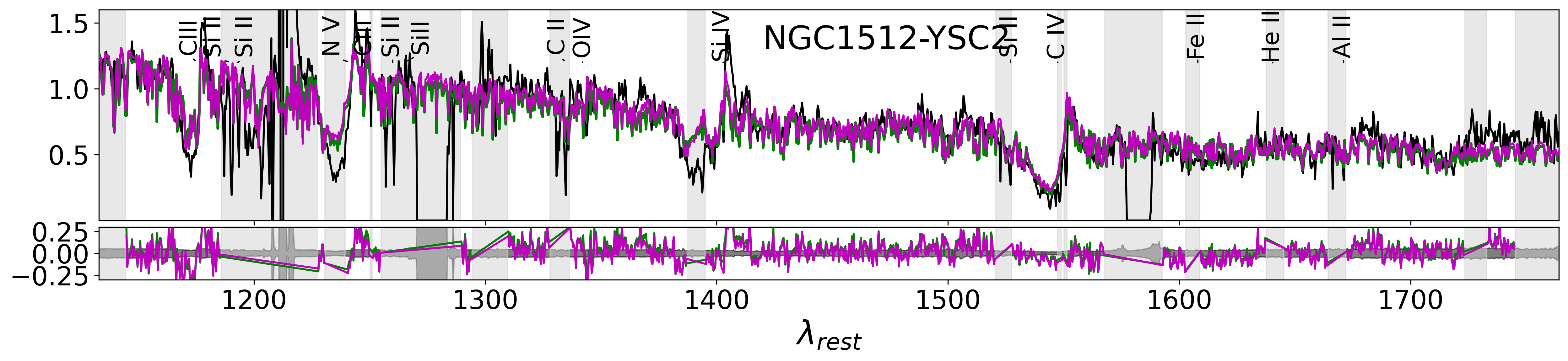}
	      
    \caption{Fits to the FUV spectra of the CLUES sample (black line) with the double-population model and the multi-population model (purple and green lines, respectively). Continues in the next page.}
    \label{fig:app}
\end{figure*}

\begin{figure*}
	\includegraphics[width=\textwidth]{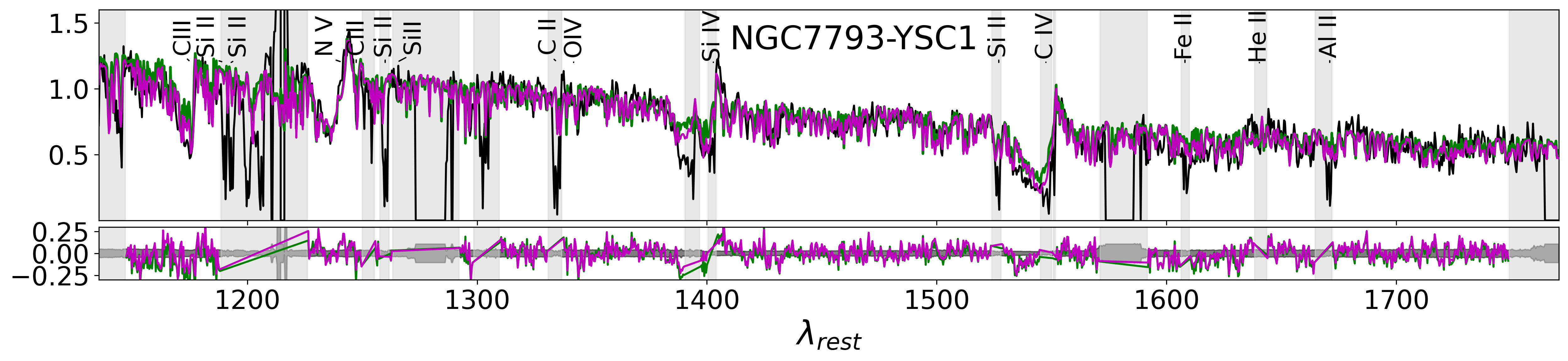}\\
	    
	\includegraphics[width=\textwidth]{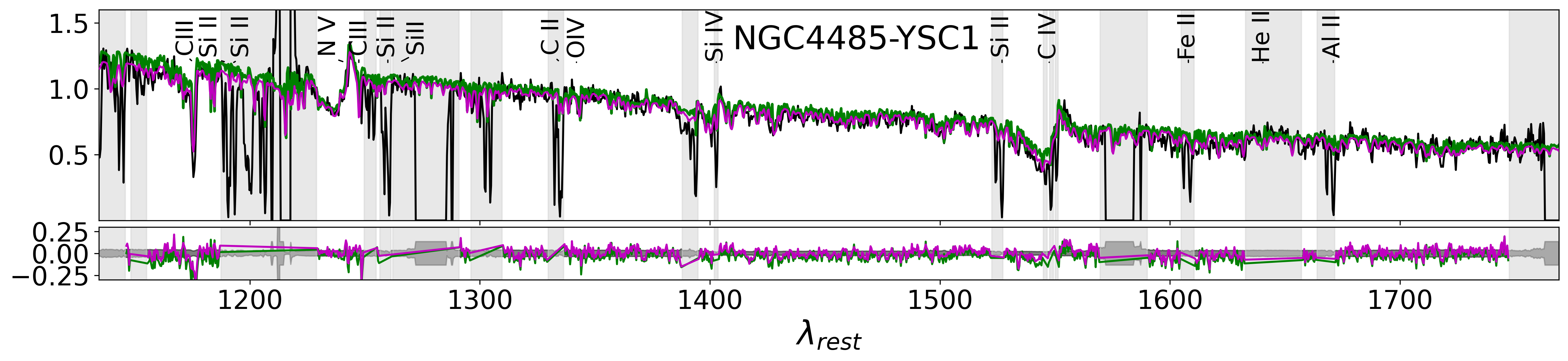}\\
	      
	\includegraphics[width=\textwidth]{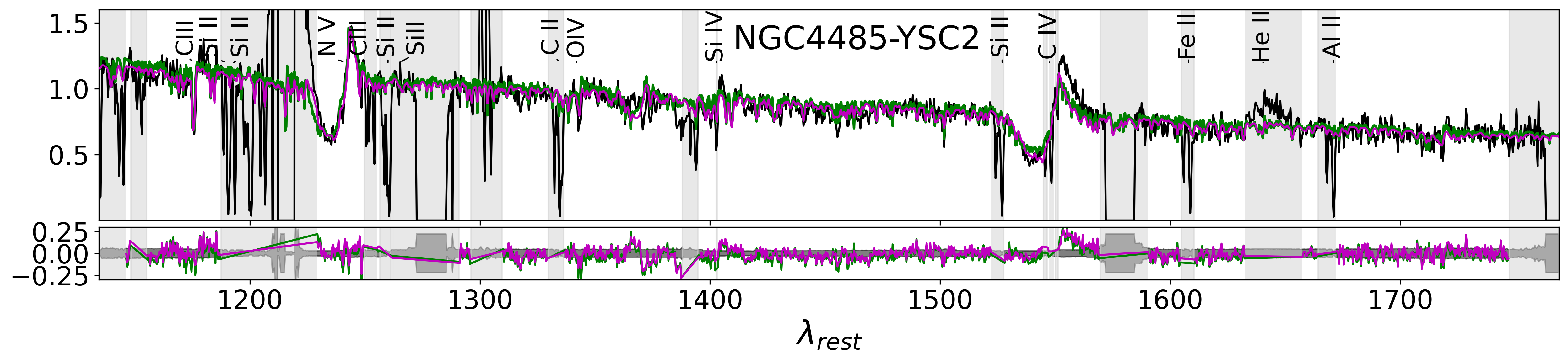}\\
	    
	\includegraphics[width=\textwidth]{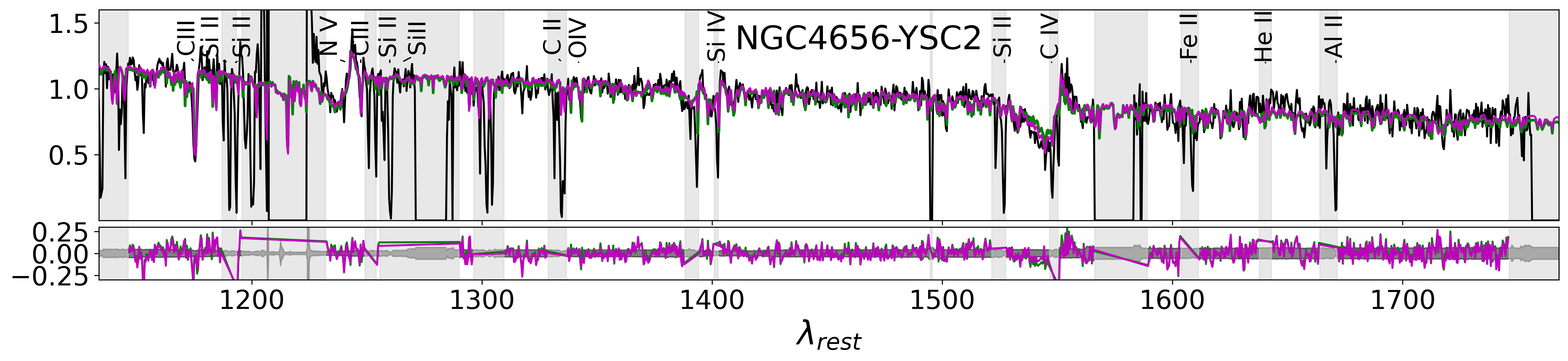}\\
	      
	\includegraphics[width=\textwidth]{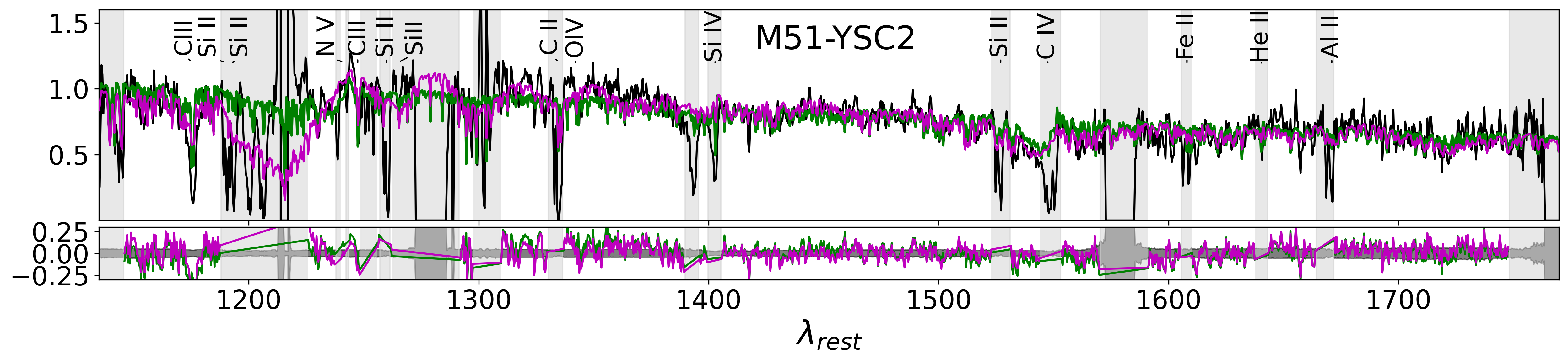}
	      
    \caption{See Figure \ref{fig:app}}
    \label{fig:app2}
\end{figure*}

\begin{figure*}
	    
	\includegraphics[width=\textwidth]{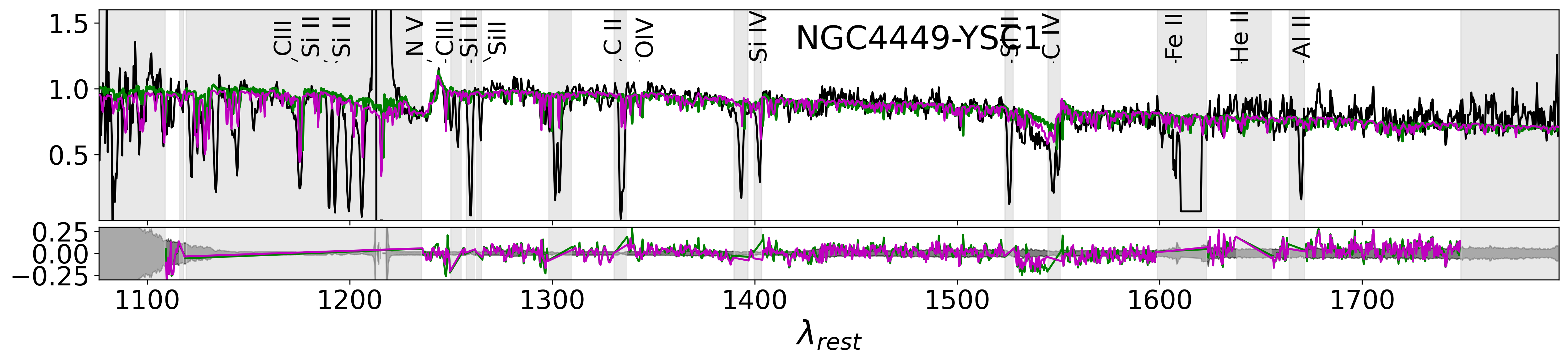}\\
	      
	\includegraphics[width=\textwidth]{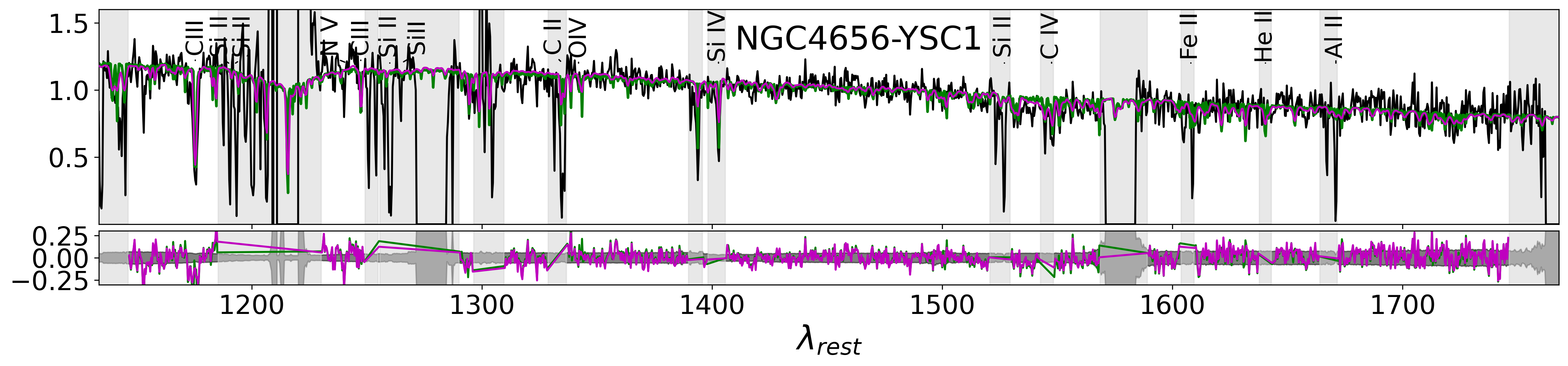}\\
	    
	\includegraphics[width=\textwidth]{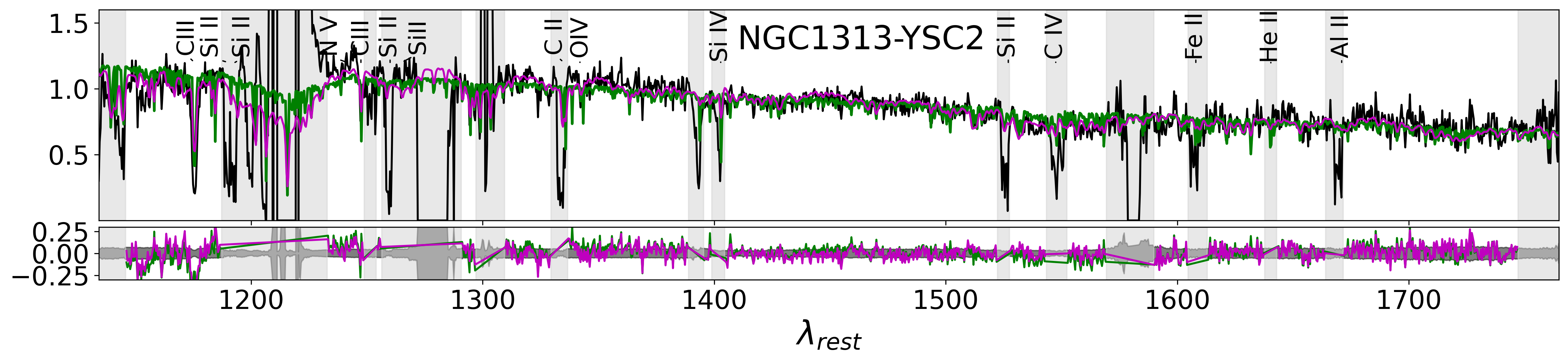}\\
	      
	\includegraphics[width=\textwidth]{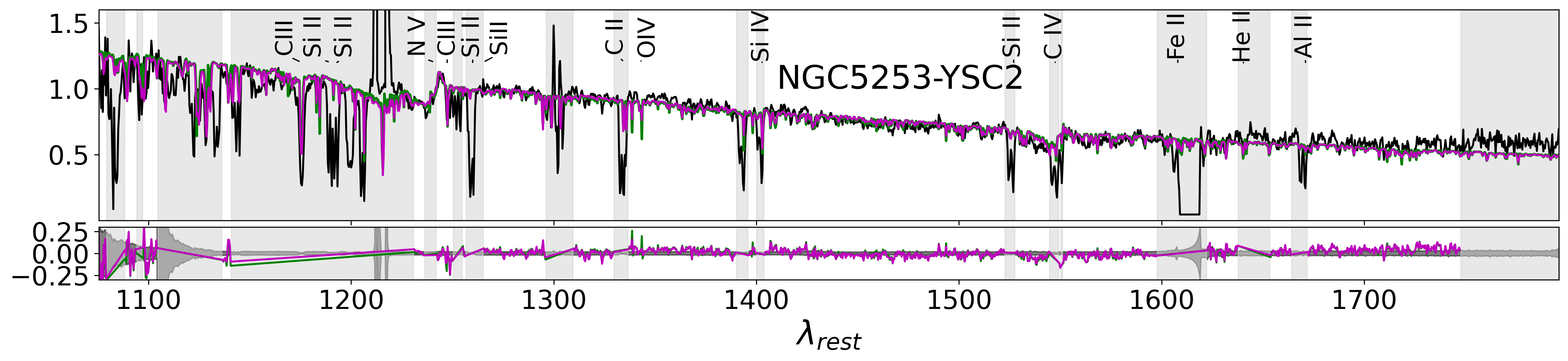}\\
	    
	\includegraphics[width=\textwidth]{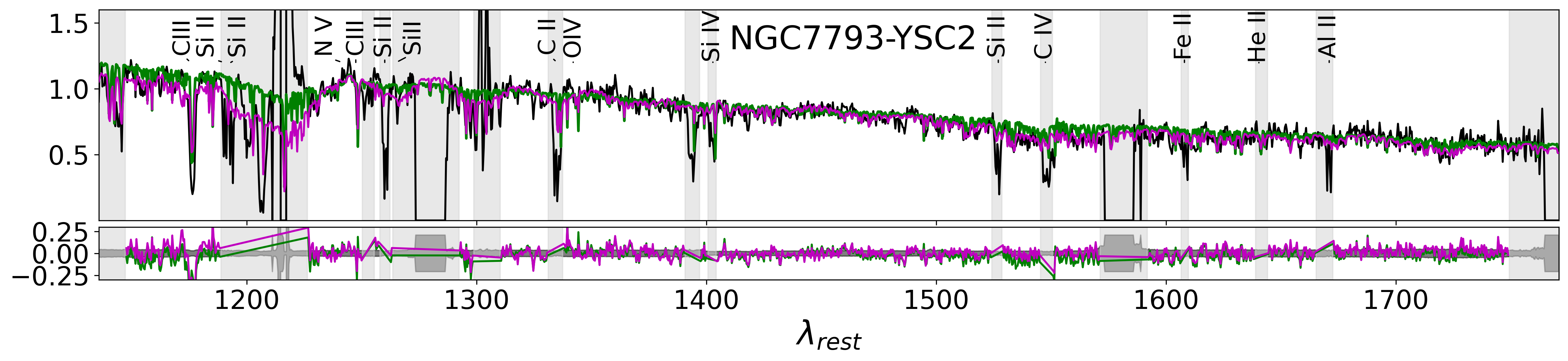}
	      
    \caption{See Figure \ref{fig:app}}
    \label{fig:app3}
\end{figure*}

\begin{figure*}

	\includegraphics[width=\textwidth]{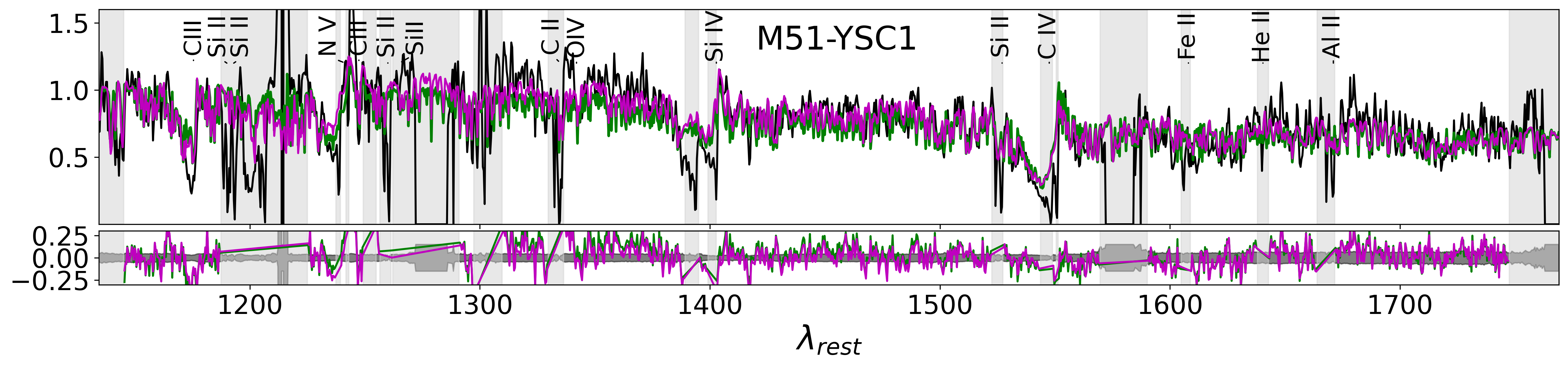}\\

	\includegraphics[width=\textwidth]{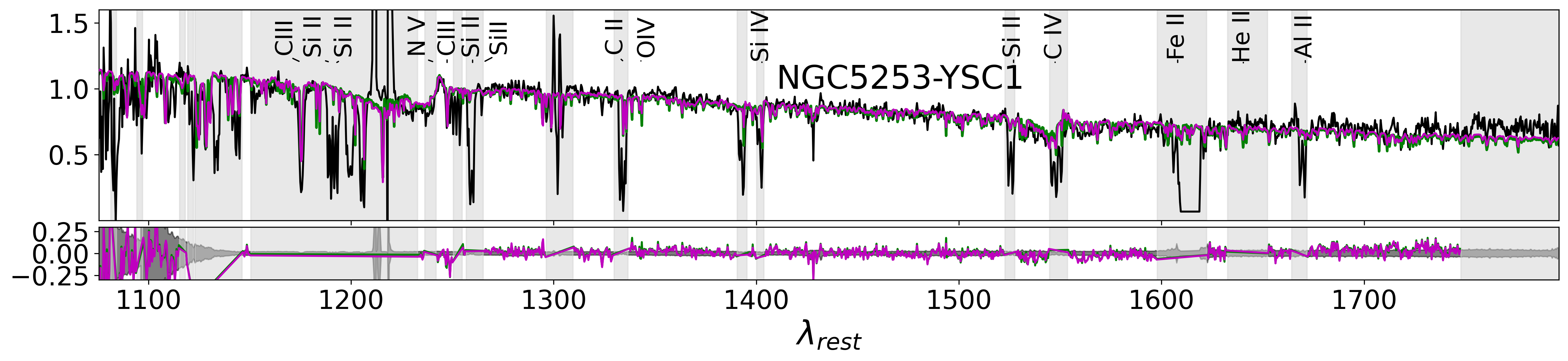}\\
	    
	\includegraphics[width=\textwidth]{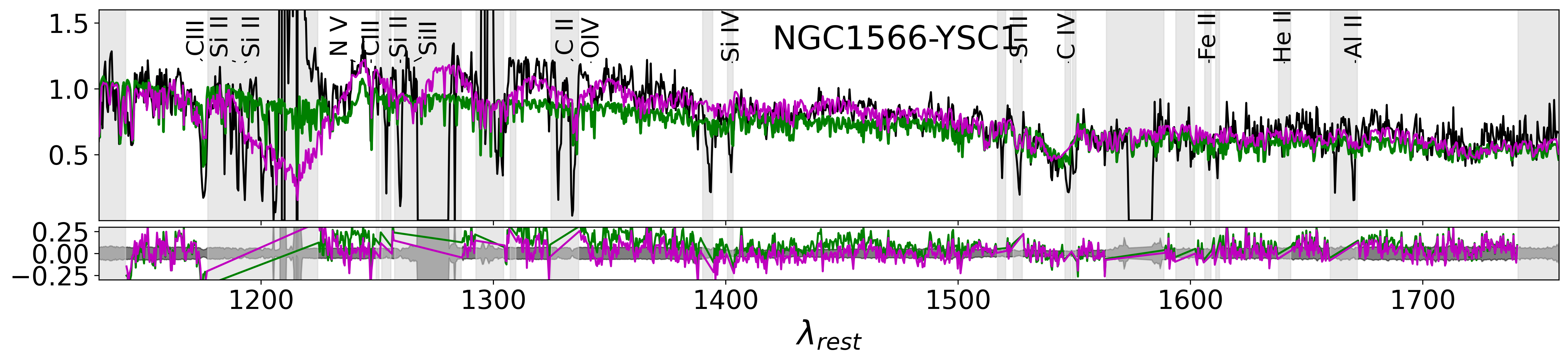}
	      
    \caption{See Figure \ref{fig:app}}
    \label{fig:app4}
\end{figure*}


\bibliography{clues1}{}
\bibliographystyle{aasjournal}



\end{document}